\documentclass[aps,prd,preprint,eqsecnum,tightenlines,floats,
groupedaddress,showpacs]{revtex4}
\usepackage{amssymb}
\usepackage{amsmath}
\usepackage{graphicx}
\usepackage{subfigure}
\usepackage{dcolumn}

\begin{document}

\date{December 4, 2024}

\title{Effects of a dark matter caustic passing through the Oort cloud}

\author{Yuxin Zhao, Antonios Kyriazis and Pierre Sikivie}
\affiliation{Department of Physics, University of Florida,
Gainesville, Florida 32611, USA}

\vskip 1cm

\begin{abstract}
We investigate the effect of a dark matter caustic passing 
through the Solar System.  We find, confirming a previous 
result, that the Sun tracks the caustic surface for some 
time.  We integrate numerically the equations of motion of 
the Sun and a comet for a large number of initial conditions 
and of caustic passage properties.  We calculate the probability 
for the comet to escape the Solar System and the probability 
for it to fall within 50~A.U. of the Sun, given the initial 
semimajor axis and eccentricity of its orbit.  We find that 
the average probability for a comet to fall within 50~A.U. 
of the Sun is of order $3\times10^{-4}$ and that comets 
which are initially at a distance larger than about $10^5$~A.U. 
have a probability of order one to be ejected from the Solar 
System.

\end{abstract}

\pacs{95.33.+d}

\maketitle

\section{Introduction}

Observations imply that a large fraction, of order 27\%, of
the energy density of the Universe is some unknown substance
called ``dark matter" \cite{Bertone}. The dark matter is thought 
to be cold and collisionless.   By ``cold" it is meant that 
the primordial velocity dispersion of the dark matter fluid
is very small and may be neglected when discussing its large 
scale evolution.  By ``collisionless" it is meant that only the 
gravitational interactions of the dark matter play an important
role in its large scale evolution.  

The appearance of caustics is a generic phenomenon in the flow 
of a fluid that is both cold and collisionless.   Caustics 
are familiar in the propagation of light.  Indeed light is 
collisionless in vacuum and has small velocity dispersion 
when its source is far away.  Rainbows, the twinkling of stars, 
the shimmering of the sea, and the sharp lines of light at the 
bottom of a swimming pool on a sunny breezy day are all due to 
caustic formation in the propagation of light.  These phenomena
occur when sunlight is not diffused through scattering by
clouds.  Its velocity dispersion is then only due to the 
finite angular size of the Sun. In our parlance, sunlight 
is cold on a sunny day because, in the limit where we 
neglect the angular size of the Sun, the velocity vector of 
sunlight has a single value, or an odd number of values, at 
each point on the Earth's surface.    

Likewise, caustics are expected in the flow of cold 
collisionless dark matter.  Indeed, if cold and collisionless, the dark matter particles lie on a three-dimensional hypersurface in six-dimensional phase space.   Let us call this hypersurface the ``phase space sheet".  In the linear regime of structure formation, the sheet's location in phase space is given by the Lema\^{i}tre-Hubble law perturbed by small random peculiar 
velocities.  Thus, in the linear regime, the phase space sheet 
covers physical space only once. The dark matter has then a 
single velocity at every location in physical space. In the 
nonlinear regime, after density perturbations and their 
associated peculiar velocities have become large, the phase 
space sheet folds back upon itself in many locations and 
winds up in phase space.   It then covers physical space 
multiple times.   At any location in physical space, there 
is an odd (1, 3, 5, \dots) number of cold flows.   Caustics 
are located  at the boundaries between regions in physical 
space where the number of flows changes by two.  On one 
side of a caustic surface there are two more flows than 
on the other.  The dark matter density $d_c$ diverges at the 
caustic surface in the limit where the velocity dispersion 
of the flow forming the caustic vanishes.   In that limit, when 
approaching the caustic surface from the side with two additional 
flows
\begin{equation}
d_c(\sigma) = {A \over \sqrt{\sigma}} \Theta(\sigma)   
\label{cauden}
\end{equation}
where $\sigma$ is the distance to the caustic surface and 
$\Theta$ is the Heaviside step function.  The constant $A$ 
is a characteristic property of the caustic with dimension 
of mass/(length)$^{5 \over 2}$.   We will call $A$ the 
caustic's fold coefficient.  What is described above is 
termed a fold catastrophe ($A_2$) in mathematical language. 
It is the simplest kind of caustic in three spatial 
dimensions, and occurs on a surface.   The next simplest 
catastrophe is the cusp ($A_3$) which occurs on a line.  

There is no requirement of symmetry for caustic formation
or for their existence.  Absence of symmetry, tidal forces, 
and phase space mixing do not smear caustics.  Only velocity 
dispersion smears caustics. The primordial velocity dispersion 
of the widely accepted dark matter candidates is very small, 
much smaller than the overall velocity dispersion of galactic 
halos.  The overall velocity dispersion of the Milky Way halo 
is of order $10^{-3} c$. The primordial  velocity dispersion 
\cite{Erken} of axions is of order $10^{-17} c$, that of 
weakly interacting massive particles (WIMPs) is of order 
$10^{-12} c$ and that of sterile neutrinos is of order
$10^{-8} c$.  The velocity dispersion of a flow forming a 
caustic may be larger than the primordial value, because
the flow may have been diffused to some extent by gravitational 
scattering off inhomogeneities in the galaxy, such as globular 
clusters and molecular clouds. However, the flows that have 
fallen in and out of the galaxy only a few times, e.g. the 
fifth flow, are only slightly diffused through scattering 
by inhomogeneities in the galaxy \cite{Ipser}. They 
remain distinct and form caustics. 

Appendix A shows that present $N$-body simulations of 
structure formation, where the typical particle 
mass is $10^5M_\odot$, have inadequate resolution 
to reveal the phase space structure of galactic halos.
When the particle mass is $10^5M_\odot$, two body 
relaxation completely smears out the phase space 
structure of galactic halos, including the cold flows 
and caustics that are the topic of this paper.  Two 
body relaxation is entirely negligible for the widely 
accepted cold dark matter candidates such as axions, 
WIMPs, and sterile neutrinos.

The dark matter in galactic halos necessarily has {\it outer}
and {\it inner} caustics \cite{Arvind}.   The outer caustics 
are simple fold catastrophes located on topological spheres 
surrounding the galaxy.  They occur near where dark matter 
particles moving away from the galactic center reach their 
largest galactocentric radii before falling back in.
The inner caustics occur near where the particles with the 
most angular momentum reach their closest approach to the 
galactic center before falling back out.   The catastrophe 
structure of the inner caustics depends on the angular momentum 
distribution of the infalling particles.  If the particles fall 
in with a velocity distribution characterized by large scale 
vorticity ($\vec{\nabla} \times \vec{v} \neq 0$), the inner 
caustics are closed circular tubes whose cross section,
shown in Fig.~1, is a section of the elliptic umbilic ($D_{-4}$) 
catastrophe.  It has three cusps, one of which points away from 
the galactic center.  This type of inner caustic is called a 
``caustic ring" \cite{crdm}.   It is described in detail in 
Ref.~\cite{sing}.  If the dark matter particles have a
velocity distribution characterized by vanishing vorticity 
($\vec{\nabla} \times \vec{v} = 0$), the inner caustics have 
a catastrophe structure that is different from that of caustic 
rings.   It is described in detail in Ref.~\cite{Arvind}.

Evidence was found \cite{crdm,Duffy,Gaia} for caustic rings of 
dark matter, with radii predicted by the self-similar infall model 
of galactic halo formation \cite{FG,STW}, implying therefore that 
the dark matter falls in and out of galactic halos with large 
scale vorticity.  A description of the full phase space distribution 
of the dark matter in the Milky Way halo consistent with large scale 
vorticity and self-similarity, called the caustic ring model, is 
given in Ref.~\cite{Duffy}. Bose-Einstein condensation of cold
dark matter axions \cite{CABEC,case,Erken} as a result of their 
gravitational self-interactions has been shown to account for 
all the model properties in detail, including the appearance 
of caustic rings in the galactic plane and the pattern of 
their radii.

Dark matter caustics, whether inner or outer, are topologically 
stable and quasi-indestructible.  They are topological features in 
large scale flows and can only be destroyed by removing the 
flows in which they occur.  A dark matter caustic is no more 
affected by, say, a globular cluster passing through it than a 
rainbow is affected by a cannonball shot at it.  At most, caustics 
are only perturbed momentarily by such events.  

The outer caustics of the Milky Way halo are far from us, typically 
at hundreds of kiloparsecs from the Galactic Center. On the other hand, 
the inner caustics are nearby, at tens of kpc or less from the 
Galactic Center.   They may be near Earth, and at times move 
through the Solar System.  Caustic rings of dark matter lie in 
the Galactic plane and move outward, increasing their radii, on 
cosmological timescales.  The presence of a nearby caustic has 
dramatic implications for dark matter searches on Earth.  According 
to Ref.~\cite{Gaia}, the solar neighborhood is most likely inside 
the tricusp volume of the fifth caustic ring of dark matter in 
the Milky Way, implying that the local dark matter distribution is 
dominated by four cold flows, called ``big", ``little", ``up" 
and ``down", with known velocity vectors, and velocity dispersion 
that is less than about 70~m/s. The expected densities of the four flows, 
respectively, 20, 2, 9, and $8 \times 10^{-24}$~gr/cm$^3$, are 
much larger than the standard estimate, approximately 
$0.5 \times 10^{-24}$~gr/cm$^3$, of the local dark matter 
density derived from fits to the galactic rotation curve.
There is no disagreement here because the standard estimate
is an average on a scale of order kpc whereas the values of 
the stated local flow densities are estimates on a 10~pc 
scale.  Note in this regard that although Eq.~(\ref{cauden})
indicates that the dark matter density diverges on the 
caustic surface in the limit of vanishing velocity dispersion, 
the divergence is integrable.   Although there is a local 
overdensity, it does not imply an increase in mass over a 
large volume.  

The goal of our paper is to estimate the effects of a dark 
matter caustic passing through the Solar System.  We will 
assume the caustic to be an $A_2$ catastrophe.  The surface 
on which it occurs is assumed to be planar on the scale of 
the Solar System.  Such a caustic passage is characterized 
by the caustic fold coefficient, i.e., the $A$ value defined 
in Eq.~(\ref{cauden}), and the velocity $\vec{v}$ of the 
caustic surface relative to the Solar System.  The values 
of $A$ (of order $2 \times 10^{-3}\,\mathrm{~gr/cm^2}\, \sqrt{{\rm kpc}}$) 
and the speed $v$ (of order 1 to 10~km/s) are taken from the 
caustic ring halo model.  The effect of the passing caustic 
on the motion of planets, and the inner Solar System in 
general, is found to be very small.   The effects become 
large, however in the Solar System's  outer reaches, thought
to be inhabited by the Oort cloud.

The Oort cloud is a hypothetical reservoir of comets,
i.e., icy planetesimals, surrounding the Sun at distances
ranging from 2,000 to 200, 000~A.U.  Its existence
was postulated \cite{Opik,Oort} to explain why we
observe long period comets in spite of the fact that
they are expected to collide with the Sun, evaporate,
or be ejected from the Solar System as a result of
planetary perturbations, on relatively short timescales. 
The Oort cloud feeds comets into the inner 
Solar System when their orbits are disturbed as a 
result of stellar encounters \cite{Oort,Fernandez}  
or as a long term effect of the Galaxy's gravitational 
field \cite{galtid}.  It is thought that the Oort cloud
contains trillions of comets in a roughly spherical
distribution \cite{Weissman}.

Our paper is organized as follows.  In Sec.~II, we 
briefly describe the properties of the inner caustics 
of the Milky Way halo in the caustic ring model.  In 
Sec.~III, we discuss the gravitational field due 
to a caustic and the tidal forces it introduces into 
the Solar System.   In Sec.~IV, we obtain the motion 
of the Sun in the vicinity of a caustic surface.  In 
Sec.~V, we use analytical  methods to estimate the 
approximate size of the effects we are interested in.  
In Sec.~VI, for millions of different cases, we 
numerically integrate the equations of motion of the
Sun and a comet while a caustic surface passes by and 
derive the change in the comet's orbit.  Sec.~VII 
provides a summary and brief discussion.

\section{Dark matter caustics}

In this section, we describe the properties of the 
inner caustics of dark matter in the caustic ring model 
of the Milky Way halo.   In that model, the inner 
caustics are rings in the Galactic disk at the 
approximate Galactocentric radii
\begin{equation}
a_n \simeq {40~{\rm kpc} \over n}
\label{crr}
\end{equation}
where $n$ = 1, 2, 3, \dots.  Our own distance to the Galactic 
Center is taken to be 8.5~kpc.  The cross section of a caustic 
ring is shown in Fig.~1.  It has three cusps, one of which is 
pointing  away from the Galactic Center.  The lines between 
the cusps indicate the location, in three dimensions, of caustic 
surfaces.  The transverse sizes $p$ and $q$ of a caustic ring 
range from 1/10 to 1/100 times the ring radius $a$.

The fifth caustic ring ($n=5$) is close to us.   
Observational evidence for its existence is given in 
Ref.~\cite{Gaia} where the fifth caustic ring is described
as follows. Its radius is estimated to be $a_5 \simeq$  
8.45 kpc. The ring is not axially symmetric in that its 
transverse sizes $p$ and $q$ vary along the ring. At the 
Sun's location,  $p_5 \sim 80$~pc and $q_5 \sim 110$~pc.
We are therefore very likely within the tricusp volume of 
the fifth caustic ring.  The ring is not centered at the 
Galactic Center but approximately 700~pc to the right 
of it from our viewpoint.  

Caustic rings expand according to
\begin{equation}
a_n(t) = a_n(t_0) 
\left({t \over t_0}\right)^{{2 \over 3} + {2 \over 9 \epsilon}}
\label{rexp}
\end{equation}
where $t$ is time since the big bang, $t_0 = 13.7$~Gyrs 
is the present time, and $\epsilon$ is a parameter whose 
\textit{a priori} value lies between 0 and 1~\cite{FG}. $\epsilon$ 
is expected to have a value between 0.25 and 0.35 on 
both theoretical \cite{STW,case} and observational 
grounds \cite{crdm,Duffy}.   The tidal torque theory 
for the origin of galactic angular momentum \cite{PJE}
predicts $\epsilon = 1/3$~\cite{case}.  Adopting this 
value, we have $a_n(t) \propto t^{4 \over 3}$. Therefore 
a caustic ring centered at the Galactic Center and 
passing by us increases its distance to the Galactic 
Center at the speed
\begin{equation}
v = {4 \over 3} {8.5~{\rm kpc} \over t_0} = 0.8~{\rm km/s}~~\ .
\label{cs}
\end{equation} 
However, in the case of the fifth caustic ring, since its center
is approximately 700~pc to the right of the Galactic Center, 
the ring  moves towards the Galactic Center at a speed of 
approximately 20~km/s in a reference frame that is 
corotating with the Galaxy at our location.  In 
such a reference frame the motion of the Sun is 
approximately 11~km/s towards the Galactic Center, 
7~km/s towards the Galactic North Pole and 12~km/s 
in the direction of Galactic rotation \cite{schon}.  In
view of all this, we investigate values between 0.3~km/s 
and 30~km/s for the speed with which the Sun 
crosses a caustic  surface. 

The fold coefficients $A_n$ of the caustic ring surfaces 
in the caustic ring model were calculated in \cite{charm}.  
However, the values given there assume that the dark matter 
falls onto our Galaxy isotropically from all directions.  
According to Ref.~\cite{Banik} the infall of axion dark 
matter is not isotropic because of the formation of a large 
vortex along the Galactic rotation axis.  As a result the 
density is enhanced in  the Galactic plane.  A density 
enhancement in the Galactic plane is also necessary to 
explain the prominence of the bumps in the inner Galactic 
rotation curve attributed to  caustic rings \cite{crdm}.  
The bumps are approximately a factor 4 larger than 
the model prediction in case of isotropic infall, implying 
that the $A_n$ are increased by approximately a factor 4.  
We expect therefore the fold coefficients  of the caustic 
ring surfaces to be of order  
\begin{equation}
\{A_n: n = 1, 2, 3 ..\} \sim (1.2, 1.6, 1.6, 2, 2, ...)
\times {10^{-3}~{\rm gr} \over {\rm cm}^2 \sqrt{\rm kpc}}
\label{An}
\end{equation}
which is four times the values given in \cite{charm}.  
These $A_n$ values are meant to apply to the 
caustic ring surfaces at points which are not too 
close to the cusps, e.g., at the locations of the 
double-sided arrows in Fig.~1.   

\section{Equations of motion}

Fig.~2 depicts a caustic surface near the Solar System.
The gravitational field of the caustic, implied by 
Poisson's equation with Eq.~(\ref{cauden}) as a source, is 
\begin{equation}
\vec{g}_c(\sigma)  = 
+ 8 \pi G A \sqrt{\sigma} \Theta(\sigma) \hat{n} 
+ \vec{g}_{c0}
\label{caugrav}
\end{equation}
where $\vec{g}_{c0}$ is a constant field and $\hat{n}$ is the 
unit vector normal to the caustic surface pointing away from 
the side with two extra flows.  As in Eq.~(\ref{cauden}),
$\sigma$ is  distance to the caustic surface with the 
positive direction being the side with two extra flows, 
i.e. the direction opposite to $\hat{n}$.  The Sun moves 
under the influence of the gravitational field of the 
Galaxy as well as that of the caustic.  

In the absence of caustics, the Sun moves approximately
on a circle centered on the Galaxy with galactocentric 
radius $r_0$ and with the galactic rotation speed 
$v_{\rm rot}$.  $v_{\rm rot}$ is taken to be a constant 
220~km/s independent of galactocentric radius.   The 
effective potential per unit mass for the radial motion 
of the Sun in the absence of caustics is then
\begin{equation}
V_{\rm eff}(r_\odot) = v_{\rm rot}^2 \ln(r_\odot) + 
{l^2 \over 2 r_\odot^2}
\label{Veff}
\end{equation}
where $l = r_0 v_{\rm rot}$ is the Sun's angular momentum 
per unit mass about the Galactic Center.  Eq.~(\ref{Veff}) 
implies  that small deviations $\delta r(t) = r_\odot(t) - r_0$ 
in the radial direction obey
\begin{equation}
{d ^2 \over dt^2} \delta r = - \omega_r^2~\delta r
\label{rados}
\end{equation}
where, for $r_0 \simeq 8.5$~kpc,  
\begin{equation}
\omega_r = \sqrt{2} {v_{\rm rot} \over r_0} 
\simeq {1 \over 27~ {\rm Myr}}~~\ .
\label{radom}
\end{equation}
The Sun oscillates also in the ``vertical" direction, i.e. 
in the direction perpendicular to the Galactic plane.  The 
restoring force in this case  is the gravitational force of 
the material in the Galactic disk.  If the central density 
of that material is taken to be 
$\rho_D = 0.18 {M_\odot \over {\rm pc}^3}$ \cite{BT}, 
the angular frequency of small vertical oscillations is 
\begin{equation}
\omega_z = \sqrt{4 \pi G \rho_D} \simeq {1 \over 10~{\rm Myr}}~~\ .
\label{zom}
\end{equation}
The caustic pulls the Sun in the $- \hat{n}$ direction 
and the Sun would run away in that direction were it not 
for the opposing force of the Galaxy's gravity.   Because 
the Galactic gravitational force increases proportionately 
to distance whereas the gravitational force of the caustic 
increases only as the square root of distance, the main 
effect of the caustic on the Sun is to shift the minimum 
of its gravitational potential by an amount that depends
on $\sigma_\odot$, the distance of the Sun to the 
caustic surface.

We are only interested in the motion of the Sun in 
the direction $\hat{n}$ perpendicular to the caustic 
surface because its motion parallel to the caustic 
surface is not affected by the caustic's presence, 
at least not directly.  So, for the Sun's motion, 
we adopt the equation 
\begin{equation}
{d^2 \over dt^2} \sigma_\odot = 
- 8 \pi G A \sqrt{\sigma_\odot} \Theta(\sigma_\odot) + g_{c0} 
- \omega^2 (\sigma_\odot - \sigma_{\odot,0}^\prime)
\label{sem}
\end{equation}
where $g_{c0} = - \hat{n}\cdot\vec{g}_{c0}$.  The last 
term in Eq.~(\ref{sem}) is the harmonic restoring force 
of the Galaxy's gravity and $\sigma_{\odot,0}^\prime$ 
is the distance to the caustic surface of the minimum 
of the Sun's effective potential when the gravitational 
force due to the caustic is turned off.  We may rewrite 
Eq.~(\ref{sem}) as
\begin{equation}
{d^2 \over dt^2} \sigma_\odot =
- 8 \pi G A \sqrt{\sigma} \Theta(\sigma)
- \omega^2 (\sigma_\odot - \sigma_{\odot,0})
\label{sem2}
\end{equation}
where $\sigma_{\odot,0} = 
\sigma_{\odot,0}^\prime  + g_{c0}/\omega^2$ is the 
distance to the caustic surface of the minimum of 
the Sun's effective potential when the Sun is on the 
side of the caustic surface without the two extra 
flows.

The double arrows in Fig.~1 show different ways in 
which the Sun may cross the surface of a caustic ring.
If the Sun enters or leaves the tricusp volume by 
passing through the vertical curved surface on the 
left of the figure, the relevant value of $\omega$ 
is $\omega_r$ given in Eq.~(\ref{radom}) since 
$\hat{n}$ points radially towards the Galactic 
Center in this case. If the Sun crosses one of the 
two other, outward slanted, surfaces of the caustic 
ring, the appropriate value of $\omega$ is intermediate 
between the values given in  Eqs.~(\ref{radom}) and 
(\ref{zom}).  In our numerical work, we will mostly 
set $\omega = \omega_r$ but also try larger values.

The equation of motion for a Solar System object 
other than the Sun itself - a planet, asteroid 
or comet - is taken to be 
\begin{equation}
{d^2 \over dt^2} \vec{x} = \hat{n} 
[8 \pi G A \sqrt{\sigma(\vec{x})} 
\Theta(\sigma(\vec{x}))
+ \omega^2 (\sigma(\vec{x}) - \sigma_{\odot,0})]
+ {G M_\odot \over r^3} (\vec{x}_\odot - \vec{x})
\label{cem}
\end{equation}
where $\vec{x}$ is the position of the object, 
$\sigma(\vec{x})$ is the corresponding distance to 
the caustic surface, $\vec{x}_\odot$ is the position 
of the Sun, and $r = |\vec{x} - \vec{x}_\odot|$.  
The equation of motion~(\ref{cem}) is with respect
to a reference frame that is corotating with the 
Galaxy at the Sun's location.  It includes the 
gravitational forces on the object from the caustic, 
the Galaxy and the Sun.  It ignores the much smaller 
Coriolis force in that reference frame and the 
gravitational forces due to other Solar System 
objects.  In Sec.~IV, we numerically integrate
Eq.~(\ref{sem2}) by itself.  In Sec.~VI, we 
numerically integrate Eqs.~(\ref{sem2}) and 
(\ref{cem}) simultaneously.  

\section{The motion of the Sun}

In this section we discuss the effect of a passing caustic on 
the motion of the Sun relative to the Galaxy.  Many of the results 
here were already obtained in refs.~\cite{natarajan,sankha}.  We 
rederive and elaborate them because of their relevance to the main 
topic of this paper, i.e. the effects of the passing caustic on 
the Oort cloud.

We restrict ourselves to the Solar motion perpendicular 
to the caustic surface since its motion parallel to the 
caustic surface  is not directly modified by the caustic's  
presence.  Let $x$ be the coordinate distance with respect 
to the Galaxy in the $-\hat{n}$ direction, i.e., 
perpendicular to the caustic surface and with 
positive $x$ in the direction with  two extra dark 
matter flows.  We  choose the origins of time $t$ and 
space $x$  so that the position of the caustic surface 
is given by
\begin{equation}
x_c(t) = v_c t~~\ .
\label{cx}
\end{equation}
$v_c$  is the velocity, assumed constant,  of the caustic
surface with respect to the Galaxy.  The equation of motion
of the  Sun is
\begin{equation}
{d^2 x_\odot \over dt^2}  =
- 8 \pi G A \sqrt{x_\odot(t) - v_ct} \Theta(x_\odot(t) - v_c t)
- \omega^2 (x_\odot(t) - x_0)
\label{sem3}
\end{equation}
obtained from Eq.~(\ref{sem2}) by substituting
$\sigma_\odot(t) = x_\odot(t) - v_c t$.  $x_0$ is
the equilibrium position of the Sun when it is on
the side of the caustic surface without the two
extra flows.

Let us suppose at first that the caustic surface is at
a fixed position $x_c$, with $v_c = 0$.  The equilibrium
 position of the Sun is then at $x_{\rm eq}$ such that
\begin{equation}
8 \pi G A \sqrt{x_{\rm eq} - x_c} \Theta(x_{\rm eq} - x_c)
+ \omega^2 (x_{\rm eq} - x_0) = 0~~\ .
\label{eqx}
\end{equation}
One readily finds that
\begin{equation}  \label{eqs}
  \begin{aligned}
    x_{\rm eq} &= x_0  &&  \text{  when } x_c > x_0\\
    &= x_0 + D - \sqrt{D(D + 2 (x_0 - x_c))} &&  \text{  when } x_c < x_0 ~~\ ,
  \end{aligned}
\end{equation}
where
\begin{equation}
D = {1 \over 2}\left({8 \pi G A \over \omega^2}\right)^2
= 94~{\rm pc}
\left({A \over 2\times 10^{-3} {\rm gr/cm^2\sqrt{kpc}}}\right)^2
\left({1/20~{\rm Myr} \over \omega}\right)^4~~\ .
\label{D}
\end{equation}
When $x_c$ approaches $x_0$ from below
\begin{equation}
x_{\rm eq} = x_c + {(x_o - x_c)^2 \over 2 D} 
+ {\cal O}((x_0-x_c)^3/D^2)~~\ ,
\label{lim}
\end{equation}
i.e. the equilibrium position tracks the position of 
the caustic.  Fig.~3a shows the  equilibrium position 
$x_{\rm eq}(t)$ as a function of time when the caustic 
moves according to Eq.~(\ref{cx}).  The graph has a 
kink at time $t_0 = x_0/v_c$ as  the equilibrium position 
abruptly stops at $x_0$ whereas the caustic passes by 
$x_0$ with constant velocity $v_c$.  

Fig.~3b shows the motion of the Sun in case the Sun is
initially on the side of the caustic surface with two 
extra flows at its equilibrium position and moving
with the same velocity as its equilibrium position. 
The motion was obtained by numerically solving
Eq.~(\ref{sem3}).  For $t<t_0$, the  Sun closely 
tracks its  equilibrium position, as expected from 
adiabaticity.  As $t$ approaches $t_0$, the Sun 
therefore closely tracks the position of the caustic.  
For $t>t_0$, the Sun oscillates about $x_0$ with 
angular frequency $\omega$.  There is no kink in 
the graph of $x_\odot(t)$. Since the Sun has the 
same velocity as the caustic at time $t_0$, the 
amplitude of its oscillation about $x_0$ after 
time $t_0$ is 
\begin{equation}
\Delta x = {v_c \over \omega} = 
20.4~{\rm pc} \left({v_c \over {\rm km/s}}\right)
\left({1/{\rm 20~Myr} \over \omega}\right)~~\ .
\label{delx}
\end{equation}
Fig.~3c shows the motion of the Sun in case the Sun 
is initially at rest at its equilibrium position 
on the side of the caustic without the two extra 
flows and the caustic moves toward the Sun, with 
$v_c<0$.  In this case, the Sun stays at $x_0$ 
untill $t_0$ and oscillates with amplitude $\Delta x$ 
about its equilibrium position when  $t>t_0$.  
Fig.~3d shows a general case with $v_c > 0$ where 
the Sun oscillates about its equilibrium position 
both before and after $t_0$.  In the case depicted 
by Fig.~3b, the Sun crosses the caustic surface
only once but tracks it for a long time, during 
which  the relative velocity of the Sun and 
caustic is very small.   In the case depicted 
by Fig.~3c, the Sun crosses the caustic surface
three times but with relatively larger velocities.
In the case depicted by Fig.~3d, the Sun crosses
the caustic surface five times, with still larger 
velocities for the first three crossings.

We note that when caustic rings of dark matter 
plow through galactic disks on cosmological timescales, 
they increase the velocity dispersion of 
the stars in the disks through the gravitational 
drag and release effect described above.  The 
observed velocity dispersion of stars in galactic 
disks~\cite{BT,Skuljan,Gaia2,Gaia3,Lian} may 
be due to  this effect, at least in part.

\section{Analytical estimates of the effects on the Oort Cloud}

In this section we give analytical estimates 
of the effects we are interested in, mainly the 
probability for a comet to escape the Solar System
and the probability for it to fall within 50~A.U. 
of the Sun.  Although only approximate, the 
analytical estimates explain the orders of 
magnitude of the numerical results obtained 
in Sec. VI, and provide intuition for how 
the passage of a caustic disturbs the orbits 
of comets.

In the Cartesian reference frame shown in Fig.~2, 
where the Sun is at rest, the equation of motion 
of a Solar System object is 
\begin{equation}
{d^2 \over dt^2} \vec{r} = \hat{n}
\{ 8 \pi G A [\sqrt{\sigma(\vec{r})} \Theta(\sigma(\vec{r}))
- \sqrt{\sigma_\odot} \Theta(\sigma_\odot)]
+ \omega^2 (\sigma(\vec{r}) - \sigma_\odot) \}
- {G M_\odot \over r^3} \vec{r}
\label{cem2}
\end{equation}
obtained from Eq.~(\ref{cem}) by subtracting the 
acceleration of the Sun.   Fig.~4  shows the strengths 
of the gravitational field of the Sun, the tidal field 
of the caustic and the tidal field of the Galaxy as a 
function of distance $\sigma$ to the caustic surface.  
In that figure, the Sun is assumed to be on the caustic 
surface, the fold coefficient of the caustic is taken 
to be $A = 2\times 10^{-3}~$gr/cm$^2 \sqrt{\rm kpc}$ 
and $\omega = \omega_r$.  The figure shows that
for $\omega = \omega_r$  the tidal field of the 
Galaxy exceeds the gravitational attraction of 
the Sun at distances from the Sun larger than
$3\times 10^5$~A.U.  For $\omega = \omega_z$ that
cross-over distance is $1.6 \times 10^5$~A.U.  The 
cross-over distance is approximately the radius 
of the Hill sphere of the Sun.  Objects outside the 
Hill sphere are ejected from the Solar System by 
the tidal field of the Galaxy \cite{Antonov,Heisler}.  
Fig.~4 also shows that, when the Sun is on the caustic 
surface, the tidal force from the caustic exceeds the 
gravitational pull of the Sun for distances to the Sun 
of order $10^5$~A.U. or larger.
    
Eq.~(\ref{cem2}) shows that the dimensionless 
parameter that gives the strength of the caustic 
tidal field relative to the gravitational field 
of the Sun is 
\begin{equation}
\zeta = 8 \pi {A a^{5 \over 2} \over M_\odot} 
= 1.24 \left({a \over 10^5~{\rm A.U.}}\right)^{5 \over 2}
\left({ A \over 2 \times 10^{-3}~{\rm gr/cm}^2~\sqrt{\rm kpc}}\right)
\label{zeta}
\end{equation}
where $a$ is the semimajor axis of the object's 
orbit.  $\zeta$ is very small for the planets, of order 
$3 \times 10^{-13}$ for the Earth.  Although the effect 
of a caustic on the Earth and the other planets may 
be observable with present day astrometry, we do 
not pursue this idea here largely because there is 
no reason to think that there is a caustic in our 
vicinity at the moment.  The passage of a caustic 
surface through the Solar System is a relatively 
rare event.  Indeed, since the distance between caustic 
rings near $n=5$ is of order kpc and the rings move 
outward at a speed of order km/s~=~1.02~pc/Myr, the 
typical time interval between the passing of successive 
caustic rings is of order Gyr.  When a caustic ring 
is nearby, the Sun may cross its surface several 
times at shorter intervals.  If we take the distance 
between surfaces of a caustic ring to be of order 
100 pc, which is appropriate for $n=5$, and the Sun 
travels between them at speeds of order 10 km/s, the 
time interval between crossings is of order 10 Myr, 
which is still very long compared to the historical 
record.

We do expect large effects when $a \simeq 10^5$~A.U. 
Let us assume here, for the purpose of making rough 
estimates, that the caustic surface  passes through 
the Solar System with constant velocity  $\vec{v} = 
v \hat{n}$. A more accurate description of the motion 
of the Sun in the vicinity of a caustic surface was 
given in Sec. IV.  The time it takes the caustic 
surface to cross the orbit of a Solar System object 
is then
\begin{equation}
t_{\rm cross} = {2 a \over v} = 0.95~{\rm Myr} 
\left({a \over 10^5~{\rm A.U.}}\right)
\left({{\rm km/s} \over v}\right)
\label{cross}
\end{equation}
whereas the object's orbital period is
\begin{equation}
\tau = 31.6~{\rm Myr}
\left({a \over 10^5~{\rm A.U.}}\right)^{3 \over 2}~~\ .
\label{period}
\end{equation}
Whenever the caustic has a sizeable  effect, i.e., whenever 
$\zeta$ is not minuscule, the orbit crossing time is short
compared to the period.  The change in velocity caused by 
the caustic passage is
\begin{equation}
\Delta\vec{v} = - 8 \pi G A \hat{n} \int_{-\infty}^{+ \infty}
dt ~[\sqrt{\sigma_\odot(t)} \Theta(\sigma_\odot(t))
- \sqrt{\sigma(t)} \Theta(\sigma(t))]
\label{imp}
\end{equation}
with
\begin{equation}
\sigma_\odot(t) = v (t - t_\odot) ~~~~{\rm and}~~~
\sigma(t) = \sigma_\odot(t) - \hat{n}\cdot\vec{r}(t)
\label{sigmas}
\end{equation}
where $\omega = {2 \pi \over \tau}$ and $t_\odot$ is the time
when the caustic surface passes the Sun.

A rough estimate of the impulse given by the caustic to 
a comet is obtained by  assuming that the comet is on, and 
keeps moving  on, a circular orbit of radius $a$ during 
the caustic passage, i.e.
\begin{equation}
\vec{r}(t) = a \{ \cos [\phi_0 + \omega (t - t_\odot)]~\hat{x}
+ \sin [\phi_0 + \omega (t - t_\odot)]~\hat{y} \} ~~\ .        
\label{vecr}  
\end{equation}
In the coordinate system defined by Fig.~2, let 
\begin{equation}
\hat{n}  = \sin \theta ~\cos \phi_c~\hat{x}
+ \sin \theta ~\sin \phi_c~ \hat{y}
+ \cos \theta ~ \hat{z}~~\ .
\label{hatn}
\end{equation}
In terms of $\phi_0$ defined by Eq.~(\ref{vecr}) and  
$\theta$  and $\phi_c$ defined by Eq.~(\ref{hatn}),  
Eq.~(\ref{imp}) becomes
\begin{eqnarray}
\Delta \vec{v} &\simeq& - 8 \pi G A \hat{n}
\Big[ \int_{t_\odot}^{+ \infty} dt \sqrt{v (t-t_\odot)}
\nonumber\\
&-& \int_{-\infty}^{+\infty} dt 
\sqrt{v(t-t_\odot) - a \sin\theta \cos(\omega t - \delta)}~
\Theta(v(t-t_\odot) - a \sin\theta \cos(\omega t - \delta)) \Big]
\label{imp2}
\end{eqnarray}
where 
\begin{equation}
\delta = \omega t_\odot + \phi_c - \phi_0~~\ .
\label{phase}
\end{equation}
The two integrals in Eq.~(\ref{imp2}) diverge at 
large $t$ but the divergences cancel each other.  
To estimate the RHS of Eq.~(\ref{imp2}) we expand the 
square root in the second integral to first order in 
$a/v(t-t_\odot)$ and neglect the difference  between 
the time $t_\odot$ when the caustic surface passes 
the Sun and the time it passes the comet.  This yields
\begin{equation}
\Delta \vec{v}  \simeq  - 4 \pi G A a \sin \theta ~\hat{n}~
 \int_{t_\odot}^{+ \infty} dt 
{\cos (\omega t - \delta) \over \sqrt{ v(t - t_\odot)}}~~\ .
\label{imp3}
\end{equation}
Analytical results for the integrals in Eq.~(\ref{imp3}) 
are known.  Substituting them yields
\begin{equation}
\Delta \vec{v} \simeq -
\zeta \hat{n} \sqrt{\pi v_0^3 \over 2 v} {1 \over 2}
\sin \theta [\cos(\phi_c - \phi_0) + \sin (\phi_c - \phi_0)]
\label{imp4}
\end{equation}
in terms of the speed  
\begin{equation}
v_0 = {2 \pi a \over \tau} = \sqrt{G M_\odot \over a}
\label{vnot}
\end{equation}
of the comet's initial orbit.

\subsubsection{Energy transfer}

Whether a comet is ejected from the Solar System depends
on the amount of energy it exchanges with the caustic.   
Whereas initially it has energy per unit mass 
\begin{equation}
E_{\rm in} = - {1 \over 2} \vec{v}_0\cdot\vec{v}_0
\label{inen}
\end{equation}
the final value is 
\begin{equation}
E_{\rm fin} \simeq - {1\over 2} \vec{v}_0\cdot\vec{v}_0
+ \vec{v}_0\cdot\Delta\vec{v}
+ {1 \over 2} \Delta \vec{v} \cdot \Delta \vec{v}~~\ .
\label{finen}
\end{equation}
Using Eq.~(\ref{finen}) we may estimate the average 
final energy for given $\theta$, $A$, $v$ and $a$, the 
average being taken over all possible phases $\phi_0$.   
Since $\langle \Delta \vec{v} \rangle = 0$,
\begin{equation} 
\langle E_{\rm fin} \rangle = 
- {1\over 2} v_0^2
+ {1 \over 2} \langle \Delta \vec{v} \cdot \Delta \vec{v} \rangle
= {G M_\odot \over 2 a}
( -1 +\zeta^2 {\pi v_0 \over 8 v} \sin^2 \theta)~~\ .
\label{avfinen}
\end{equation}
Substituting Eq.~(\ref{zeta}) and 
\begin{equation}
v_0 = 95~{\rm{m \over s}}\left({10^5~{\rm A.U.} \over a}\right)^{1 \over 2}
\label{v0}
\end{equation}
we find that $\langle E_{\rm fin} \rangle > 0$ for 
\begin{equation}
a \gtrsim 2 \times 10^5~{\rm A.U.}
\left({1 \over \sin \theta}\right)^{4 \over 9} 
\left({ v \over {\rm km/s}} \right)^{2 \over 9}
\left({ 2\times 10^{-3}~{\rm gr/cm^2 \sqrt{kpc}} \over A} \right)^{4 \over 9}
~~\ .
\label{aej}
\end{equation}
We expect therefore that comets at a distance of order
$2 \times 10^5$ A.U. or more have a large probability 
to be ejected from the Solar System, assuming $\sin \theta$
is of order one.  When $\sin\theta = 0$ the caustic passing 
has no effect since the caustic's tidal field vanishes when 
the caustic surface is parallel to the comet's orbital plane.

In Sec. VI, the orbits of comets are obtained numerically 
in millions of cases, and the probability of ejection from the 
Solar System calculated as a function of the initial orbit 
semimajor axis $a$ and  eccentricity $\epsilon$, and various 
variables characterizing the caustic passage.  Eq.~(\ref{aej}), 
which assumes circular orbits, is qualitatively consistent with 
the numerical results of Sec. VI as it is found there that 
for $\epsilon = 0$ the ejection probability is of order 50\% 
when $a =2 \times 10^5$~A.U. and $\theta = 90^\circ$.  
    
We may also see from Eq.~(\ref{finen}) that a comet on 
a circular orbit cannot be ejected from the Solar System 
unless its orbital radius $a$ is above some critical 
value $a_c$.  In the impulse approximation, the most 
favorable case for ejection is $\Delta \vec{v}$ in the 
direction of $\vec{v}_0$, i.e.  $\theta = 90^\circ$ and 
$\phi_c - \phi_0 = 90^\circ$.  Furthermore  Eq.~(\ref{finen}) 
implies that the smallest $\Delta \vec{v}$ necessary for 
ejection is $\Delta \vec{v} = (\sqrt{2} - 1) \vec{v}_0$.  
Using this and Eqs.~(\ref{zeta}), (\ref{imp4}) and 
(\ref{vnot}), we find
\begin{equation}
a_c = 1.3 \times 10^5~{\rm A.U.} 
\left({ v \over {\rm km/s}} \right)^{2 \over 9}
\left({ 2\times 10^{-3}~{\rm gr/cm^2 \sqrt{kpc}} \over A} 
\right)^{4 \over 9}~~\ .
\label{ac}  
\end{equation}
This agrees within 30\%  with the more accurate and more 
general results of Sec. VI.

\subsubsection{Angular momentum transfer}

Whether a comet is caused by a  caustic passage to
fall within a small distance $d$ (say 50~A.U.) from the
Sun depends mainly on the accompanying change in the 
comet's angular momentum.  In Keplerian motion, the 
shortest distance of the approach to the Sun is
\begin{equation}
r_{\rm min} = {\ell^2 \over GM_\odot + \sqrt{(GM_\odot)^2 + 2 E \ell^2}}
\label{rmin}
\end{equation}
where $\ell$ is the orbiting object's angular momentum 
magnitude per unit mass.  As before, $E$ is its energy
per unit mass.  For the highly eccentric orbits of comets
originating in the Oort cloud but passing near the Sun,
the second term under the square root in Eq.~(\ref{rmin})
is much smaller than the first and we may approximate 
$r_{\rm min}$  by
\begin{equation}
r_{\rm min} \simeq {\ell^2 \over 2 G M_\odot}~~\ .
\label{rmin2}
\end{equation}
The angular momentum per unit mass vector is
\begin{equation}
\vec{\ell}(t) = \ell_{\rm in}\hat{z}  + \Delta \vec{\ell}(t)
\label{ellv}
\end{equation}
where $\ell_{\rm in}$ is its  initial magnitude and
\begin{equation}
\Delta \vec{\ell}(t) = 
\int_{- \infty}^t dt^\prime ~\vec{\gamma}(t^\prime)
\label{ellch}
\end{equation}
is the change caused by the passing caustic's torque
\begin{equation}
\vec{\gamma}(t) = \vec{r}(t) \times \hat{n} (- 8 \pi G A)
[\sqrt{\sigma_\odot(t)} \Theta(\sigma_\odot(t)) 
- \sqrt{\sigma(t)} \Theta(\sigma(t))]~~\ .
\label{torq}
\end{equation}
One can readily show that, when the Solar System is 
on the side of the caustic surface with two extra 
flows, the torque $\vec{\gamma}$ causes a long term 
precession of $\vec{\ell}$ around $\vec{n}$, unless 
$\theta = 0$ or $\pi/2$.

For a comet to fall within a distance $d$ of the Sun, 
it is necessary that 
\begin{equation}
\ell^2 = (\ell_{\rm in} - \Delta l_z)^2 + 
(\Delta \ell_x)^2 + (\Delta \ell_y)^2 < 2 G M_\odot  d~~\ .
\label{ellcond}
\end{equation} 
Because the LHS of this inequality is a sum of squares, 
each term must be smaller than the RHS.  This results in
two conditions: (1) the torque must cause $\ell_z$ to be 
driven to near zero, and (2) the torque  may generate 
only very small  $\Delta \ell_x$ and  $\Delta \ell_y$.   
We may obtain a rough idea of what the two conditions imply
by setting
\begin{equation}
\Delta \vec{\ell} = \vec{r_0} \times \Delta \vec {v}
\label{rdl}
\end{equation}
where $\vec{r}_0$ is the  average position of the comet
while the caustic crossed its orbit and $\Delta \vec{v}$
is the impulse per unit mass imparted to it.  For 
$\Delta \vec{v}$ we use 
\begin{equation}
\Delta \vec{v} \sim \pm 
\zeta \hat{n} \sqrt{\pi v_0^3 \over 2 v} {1 \over 2}
\sin \theta 
\label{imp5}
\end{equation}
which is a typical value in view of Eq.~(\ref{imp4}).   
The two aforementioned conditions are then qualitatively  
\begin{equation}
|r_0 (v_0 - \Delta v \sin \theta \sin (\phi_c - \phi_0)) | 
\lesssim  \sqrt{2 G M_\odot d}
\label{cond1}
\end{equation}
and
\begin{equation}
r_0 ~\Delta v~\cos \theta  \lesssim \sqrt{2 G M_\odot d} ~~\ .
\label{cond2}
\end{equation} 
Since $r_0 v_0 >> \sqrt{2 G M_\odot d}$, the first condition 
requires $\Delta v $ to be of order $v_0$ or larger.  The 
second condition requires then that $\theta$ be very close 
to 90$^\circ$.  Setting $r_0 \sim a$ and substituting 
Eqs.~(\ref{imp5}) and (\ref{zeta}), we find 
that 
\begin{equation}
\Big|\theta - {\pi \over 2}\Big| < 0.05 
\left({d \over 50~{\rm A.U.}}\right)^{1 \over 2}
\left({v \over {\rm km/s}}\right)^{1 \over 2}
\left({10^5~{\rm A.U.} \over a}\right)^{11 \over 4}
\left({2 \times 10^{-3}~{\rm gr/cm^2 \sqrt{kpc}} \over A}\right)
\label{cond2v2}
\end{equation} 
is required by the second condition.  Only comets whose initial 
orbital plane is very nearly perpendicular to the caustic surface, 
to  within approximately $3^\circ$ for the nominal parameter values 
on the RHS of Eq.~(\ref{cond2v2}), can be caused to fall to within 
50~A.U. of the Sun. 

The first condition requires that $\Delta v \gtrsim v_0$  and hence 
that $a$ be larger than a critical value.  That critical value is 
approximately the same as $a_c$ given in Eq.~(\ref{ac}) since ejection 
from the Solar System also requires $\Delta v \gtrsim v_0$. 

The conclusions in this subsubsection are qualitatively consistent  
with the numerical simulations reported in Sec.~VI.

\section{Numerical study of the effects on the Oort Cloud}

In this section, we report on the results of numerically 
integrating Eqs.~(\ref{sem2}) and (\ref{cem}) for the  
motion of the Sun and of a comet.  The results are reported 
in Figs.~5 to 17 which give probabilities for the comet to 
be ejected from the Solar System and probabilities for the 
comet to fall in within 50~A.U. of the Sun.

The parameters controlling the effect of a passing caustic
on a comet's orbit are as follows: 

\begin{itemize}

\item $A$ : the caustic's fold coefficient.

\item $v_c$ :  the velocity component of the caustic surface 
relative to the Galaxy in the $-\hat{n}$ direction, 
perpendicular to the caustic surface and towards the 
side with two extra flows.

\item $v_\odot$ : the initial velocity of the Sun relative
to its equilibrium point.

\item $\omega$ : the angular frequency that determines the 
restoring force of the Galaxy's gravitational potential.

\item $\theta$ and $\phi_c$:  the two angles, defined in 
Eq.~(\ref{hatn}), that give the direction of $\hat{n}$
relative to the comet's initial orbit.

\item $a$ and $\epsilon$:  the semimajor axis 
and eccentricity of the comet's initial orbit.

\item $\phi$ :  the initial position of the comet on its 
orbit.

\end{itemize}

It is of course not possible to numerically integrate 
the comet's motion for every conceivable combination 
of these parameters.   Our approach is as follows.   
First we  choose a standard case defined by  values for 
$A$, $v_c$, $v_\odot$, $\omega$, $\theta$, and $\phi_c$, 
and then deviate from it by successively changing the 
value of each one of those parameters. In each case, 
we integrate the equations of motion for a hundred values 
of $a$, a hundred values of $\epsilon$, and a hundred values 
of $\phi$, a million cases in all. We summarize the 
results by giving the probability for the comet to 
escape the Solar System and the probability for it 
to fall in within 50 A.U. of the Sun, for given $a$ 
and $\epsilon$.  

For the standard case we choose 
$A = 2 \times 10^{-3}\,{\rm gr/cm^2 \sqrt{kpc}}$, 
$v_c = 1$~km/s, and the Sun is initially at its 
equilibrium position on the side of the caustic 
surface with two  extra flows with $v_\odot = 0$, 
$\omega = 1/27$~Myr, $\theta = 90^\circ$ and 
$\phi_c = 180^\circ$.  The values of $a$ cover the 
range from $3 \times 10^3$ to $2 \times 10^5$ A.U. in 
equidistant steps.  The values of $\epsilon$ cover 
the range 0 to $1 - 10^3\,\mathrm{A.U.}/a$ in 
equidistant steps.  The upper bound on $\epsilon$ 
insures that the  comet's initial perihelion distance 
to the Sun is larger than $10^3$ A.U. The values 
of $\phi$ range from 0 to $2\pi$ with probability 
distribution
\begin{equation}
{\cal P}(\phi) d\phi = {1 \over \tau \dot{\phi}(\phi)} d\phi 
\label{prob}
\end{equation}  
where $\tau$ is the orbit period.  The initial time 
is taken to be before the Sun's distance to the 
approaching caustic surface is $2.5a$, and the  
equations of motion are integrated from this initial 
time until after the Sun has moved $2a$ past the caustic surface.  In case the Sun oscillates 
initially about its equilibrium position, the 
integration time is taken to be much longer than 
in the other cases.  The numerical integration 
is implemented using the fifth order Runge-Kutta 
method.  In the absence of caustics, the code 
conserves the numerical values of $a$ and 
$\epsilon$ to a relative accuracy of $10^{-5}$  
or better, during the whole integration 
interval.  The numerical code used in this 
project is available at \cite{code}.

Fig.~5 shows the probability for the comet to 
escape from the Solar System in the standard case
as a function of $a$ and $\epsilon$.  The escape
probabilities are indicated by the linearly scaled 
color bar on the right of the figure.  The red 
line separates the region in $(a,\epsilon)$ 
space where the probability is zero from the 
region where it is nonzero.  The figure shows
that comets with $a \gtrsim 80,000$~A.U. have 
a nonzero probability to escape and that the 
escape probability increases quickly with $a$.

Fig.~6 shows the probability for the comet to
fall within 50~A.U. of the Sun in the standard 
case as a function of $a$ and $\epsilon$.  The 
infall probabilities are indicated by a 
logarithmically scaled color bar.  Since we 
investigate only one hundred $\phi$ values, 
probabilities less than 0.01 are counted as 
zero. The figure shows that the infall probability 
can be quite large (over 10\%) in relatively 
narrow bands of $(a,\epsilon)$ space, reflecting 
the fact that the conditions for removal of angular 
momentum are somewhat special.  Generally the 
infall probability decreases with increasing 
$a$ because the comet may escape before it has 
a chance to come close to the Sun.  

Fig.~7 shows the effect of varying $\theta$ on 
the ejection probability. At $\theta$ = 0, the 
caustic has no effect since, being parallel to 
the comet's orbit, it accelerates the Sun and 
comet equally. The ejection probabilities 
decrease {\it progressively} as  $\theta$ varies 
from $90^\circ$ to zero.  Fig.~8 shows the effect 
of varying $\theta$ on  the infall probability.  
The infall probability decreases {\it abruptly} 
to zero as $\theta$ varies from $90^\circ$ to 
$86^\circ$.  As explained in Sec. V, unless 
$\theta$ is very close to $90^\circ$ [see 
Eq.~(\ref{cond2v2})] the torque due to the 
caustic's tidal field causes the comet's 
angular momentum vector to change direction, 
rather than decrease in magnitude. 

The effect of varying $\phi_c$ is shown in 
Fig.~9 for the ejection probability and in 
Fig.~10 for the infall probability.  When 
$\epsilon$ = 0, there is no change since 
$\phi_c$ loses its meaning in that limit.
The figures show a relatively strong 
dependence on $\phi_c$ when $\epsilon \neq$ 0.

The effect of having the Sun oscillate about 
its equilibrium position is shown in Fig.~11.
For the case shown, the Sun is initially on the 
side of the caustic surface with two extra flows, 
as in Fig.~3(d).  Fig.~12 shows the case where 
the Sun is initially at rest on the side of
the caustic surface without the two extra
flows, as in Fig.~3(c).  Fig.~13 shows the 
effect of changing the value of $\omega$.

Fig.~14 shows the effect of varying the 
caustic velocity $v_c$ with respect to
the Galaxy.  As might be expected, a 
slow moving caustic has more impact than 
a fast moving one. Fig.~15 shows the effect 
of varying the fold coefficient $A$.  
Fig.~16 shows the effect of dividing the 
caustic surface of the standard case into 
seven equal caustic surfaces spread over 
a thickness of 2~pc, to account for the 
possibility that the dark matter flow 
forming the caustic surface has velocity 
dispersion.

We believe it useful to give an estimate of 
the average probability for a comet in the Oort 
cloud to fall within 50~A.U. of the Sun due to 
the passage of a dark matter caustic.  Fig.~17 
shows this infall probability in the standard 
case, averaged over $\epsilon$ with a flat 
distribution and over six values of $\phi_c$ 
($0^\circ, 60^\circ, 120^\circ, 180^\circ, 
240^\circ, 300^\circ$).   
We convoluted this average probability with 
the comet number density distribution  
beyond 3000~A.U. derived from the  numerical 
simulations of the  Oort cloud reported in 
refs. \cite{Duncan,Kaib}
\begin{equation}
  n(r) \propto r^{-3.4}\quad,
\end{equation}
where $r$ is the mean radial distance of the
comet from the Sun.   This yields an average 
infall probability of 0.03 in the standard 
case which assumes $\theta = 90 ^\circ$.  We 
divide this probability by 100 because there 
is only a sizeable infall probability when 
$\theta$ is close to $90^\circ$.  We therefore 
estimate the average infall probability to be of order 
$0.01 \times 0.03 = 3 \times 10^{-4}$. The 
uncertainties are large both because the distribution  
of comets in the Oort cloud is poorly constrained and 
because the parameters characterizing a caustic passage 
have wide possible ranges.

\section{Summary and Discussion}

Caustics are a necessary property of galactic 
halos under the standard assumption that the 
dark matter is cold and collisionless. Galactic 
halos have outer caustics and inner caustics.   
Whereas the outer caustics are at hundreds of 
kiloparsecs from the Galactic Center, the inner caustics 
are at tens of kiloparsecs from the Galactic Center and 
hence much closer to us.  Generically a dark 
matter caustic is a surface at which the dark 
matter density diverges, as described by
Eq.~(\ref{cauden}).  This paper discusses 
what happens when a caustic surface passes 
through the Solar System.

The caustic is characterized by a fold coefficient 
$A$ and velocity $\vec{v}_c$ relative to a reference
frame that is comoving with the Galaxy at our 
location. Estimates for the  caustic fold 
coefficients and velocities are taken from the
caustic ring model. The caustic ring model is 
an idealized description of the phase space 
structure of the Milky Way halo. An important 
model property, especially relevant to the present 
study, is that its inner caustics are rings that 
lie in the Galactic plane. The rings expand on 
cosmological timescales.  Some of them have 
passed through the Solar System in the past, 
and others will do so in the future.  The only 
requirement, however, for the inner caustics 
to be rings in the Galactic plane is that the velocity 
field of infalling dark matter be characterized by 
large scale vorticity.  Observational evidence has 
been claimed for the model.  Furthermore, it has 
been shown that Bose-Einstein condensation of cold 
dark matter axions as a result of their gravitational 
self-interactions accounts for all the model 
properties, including the appearance of caustic 
rings in the Galactic plane and the pattern of 
their radii. A large scale merger, such as our 
expected future merger with M31, would grossly 
distort the phase space structure implied by the 
model, but there is no evidence that such a large 
scale merger occurred in the recent past. 
Fig.~1 shows the cross section 
of a caustic ring.  In the caustic ring model 
the fifth caustic ring is 
close to us.  Its radius increases approximately 
at a rate of 0.8~km/s. However,  because its 
center is displaced from the Galactic Center 
and because of the motion of the Sun relative 
to the local standard of rest, the velocity of 
the fifth caustic ring relative to the Sun is of 
order 10~km/s.

Sec.~IV discusses the motion of the Sun
in the vicinity of a caustic.  The Sun moves 
under the influence of the gravitational fields 
of the caustic and that of the Galaxy.  In the 
absence of caustic, the Sun oscillates about the 
minimum of its effective potential due to the 
Galaxy.  In the presence of the Caustic, the 
minimum of the Sun's effective potential is 
displaced.  It tracks the caustic surface
for some time, as shown in Fig.~3(a).  The 
Sun, oscillating about the minimum of its 
effective potential, also tends to track the 
caustic surface as illustrated in Figs.~3(b), 
3(c) and 3(d).

The effect of the passing caustic on a 
comet in the Oort cloud is estimated by 
analytical methods in Sec.~V and 
derived numerically for many different 
cases in Sec.~VI.  Figs.~5 through 17
give the probabilities for the comet   
to be ejected from the Solar System and 
for it to fall within 50~A.U. of the 
Sun as a function of the semimajor 
axis and eccentricity of the comet's
initial orbit.  A comet with semimajor axis larger 
than $10^5$~A.U. is likely to be 
ejected from the Solar System. Since 
the Sun is expected to have passed
through several caustic surfaces in the 
past, it is predicted that the Oort cloud 
does not extend much beyond $10^5$~A.U. The average probability for a comet to fall 
within 50~A.U. of the Sun is estimated to 
be of order $3 \times 10^{-4}$, with large 
uncertainties due to the lack of information 
on the distribution of comets in the Oort 
cloud and the wide range of parameters 
characterizing possible caustic passages.
If the Oort cloud has of order a trillion 
comets, as is commonly thought, the passage
of a dark matter caustic causes of order 
$3 \times 10^8$ comets to fall within 50~A.U. 
of the Sun over a period of order a Myr, 
i.e., approximately 300 new comets per year 
during that time.  It is predicted that 
such comet rains happened in the past 
separated by intervals of order Gyr, and 
will occur in the future.  Comet rains 
likely result in a period of increased
bombardment of the inner planets and their 
moons.

\begin{acknowledgments}

We are grateful to  Anthony Gonzalez and Lawrence Widrow
for useful comments. This work was supported in part by 
the U.S. Department of Energy under Grant No. DE-SC0022148 at 
the University of Florida.  A.K. acknowledges support by the 
Onassis Foundation - Scholarship ID: F ZS 031-1/2022-2023.

\end{acknowledgments}

\vskip 1.0cm

\appendix

\section{Resolution of $N$-body simulations of galactic halo 
formation}

\vskip8pt

In this appendix, we consider under what conditions $N$-body simulations
of galactic halo formation have sufficient resolution to describe  
the behavior of cold collisionless dark matter.  The requirement that 
the fluid is cold and collisionless places an upper limit on the mass 
of the simulated particles. We find that present $N$-body simulations 
of structure formation, where the typical particle mass is $10^5M_\odot$, 
have inadequate resolution for the stated purpose.

Consider a test particle moving with velocity $v$ through a 
fluid of particles of mass $M$ and number density $n$. For 
the sake of argument we assume the fluid to be homogeneous
and at rest.  The test particle is scattered by an angle
\begin{equation}
\delta \theta \simeq {2 G M \over b v^2}
\label{onescatt}
\end{equation}
each time it passes by a fluid particle, where $b$ is the 
impact parameter.  The number of such scatterings per unit 
time with impact parameters between $b$ and $b + db$ is 
$2 \pi~ b~ db~n~v$.   The successive scatterings are in random 
transverse directions, so that the deviation $\Delta \theta$ 
from  the original direction of motion performs a random walk 
with  step size of order $\delta \theta$.  After the test 
particle has traveled through the  fluid for a time $t$ 
\begin{eqnarray}
(\Delta \theta)^2 &=& \int_{b_{\rm min}}^{b_{\rm max}}
2 \pi~b~db~n~v~t~\left({2 G M \over b v^2}\right)^2
\nonumber\\
&=& 2.6 \times 10^{-9} \left({M \over M_\odot}\right)
\left({10^{-3}~c \over v}\right)^3
\left({t \over 10^{10}~{\rm year}}\right)
\left({\rho \over 10^{-24}~{\rm gr/cm^3}}\right)
\ln\left({b_{\rm max} \over b_{\rm min}}\right)
\label{manyscatt}
\end{eqnarray} 
where $\rho = n M$ and $b_{\rm max}$ are, respectively
the mass density and the average interparticle distance 
of the fluid, and $b_{\rm min}$ is the size of the 
fluid particles or some other short distance cutoff. 
Eq.~(\ref{manyscatt}) shows typical values of the 
relevant parameters for the case of a test particle 
traveling through the inner parts of the Milky Way 
halo.  We set 
$\ln\left({b_{\rm max} \over b_{\rm min}}\right) \simeq 10$
since much smaller or much larger values are unrealistic.
This yields
\begin{equation}
\Delta \theta \simeq 
0.05 \sqrt{M \over 10^5M_\odot}~~\ .
\label{Delt}
\end{equation}
Since the particle has traveled a distance of order
$v t \simeq 3$~Mpc, the error on its position is 
not less than
\begin{equation}
\Delta s \simeq v t \Delta \theta 
\simeq 150\,{\rm kpc} \sqrt{M \over 10^5M_\odot}~\ .
\label{Dels}
\end{equation}
Thus the error on the position is of order of the halo 
size when $M = 10^5 M_\odot$, a typical value 
in present simulations \cite{sim}.  The phase 
space structure of the halo,  including the cold 
flows and caustics  discussed in this paper, is 
then washed out.   The most widely discussed 
cold dark matter candidates are WIMPs with mass of order 
10~GeV/$c^2$, sterile neutrino with mass of 
order 3~keV/$c^2$, and axions with mass of order 
$\mu$eV/$c^2$.  Since 
$M_\odot \simeq 1.1\times 10^{57}$~GeV/$c^2$, 
the scattering of a test particle by WIMPs, 
sterile neutrinos, or axions is extremely small, 
with $\Delta \theta \lesssim 10^{-31}$.  For 
the widely discussed cold dark matter 
candidates, collisional relaxation is 
thus entirely negligible, which is why such 
particles are called collisionless. Simulated 
dark matter with $M \sim 10^5M_\odot$ is clearly
not collisionless. A reasonable criterion
for part of the phase space structure of cold 
collisionless dark matter to survive is that 
the uncertainty on particle position be less 
than 0.5~kpc.  This would require 
$M \lesssim M_\odot$ as a necessary condition
for the simulated matter to be collisionless.

\newpage

\maxdeadcycles=200

\begin{figure}
\begin{center}
\includegraphics[height=110mm]{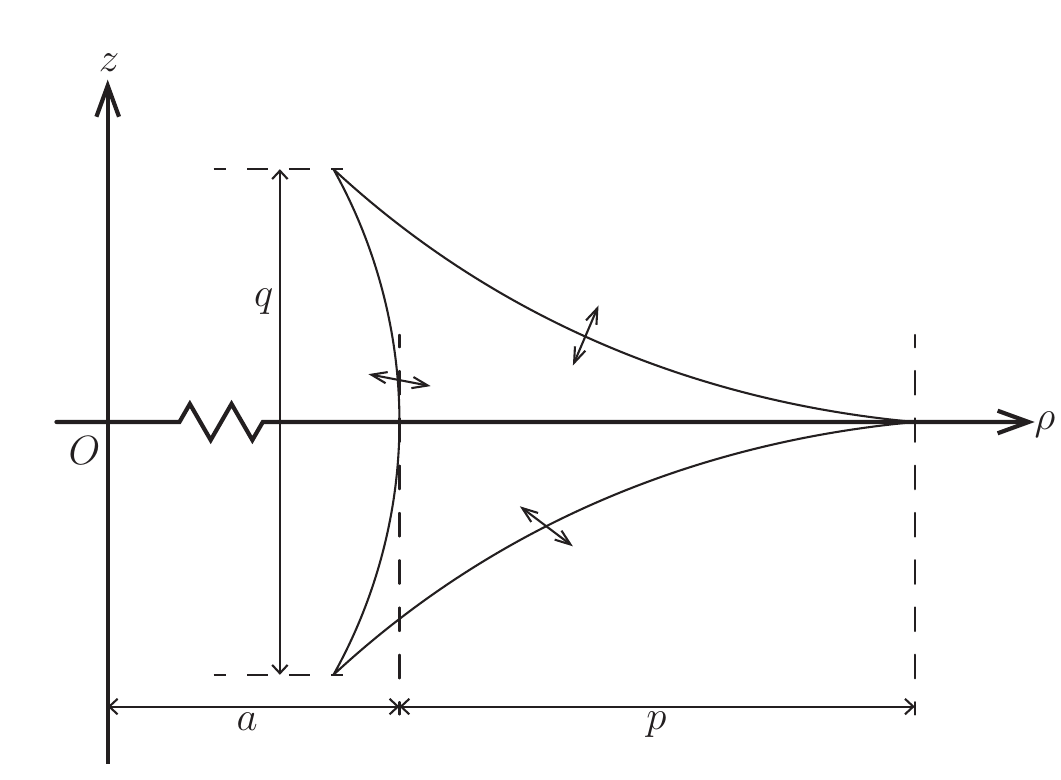}
\vspace{0.3in}
\caption{Cross section of a caustic ring of dark matter.
It has the shape of a tricusp.  The figure defines the 
caustic ring radius $a$ and transverse dimensions $p$ and 
$q$.  The Galactic Center is at $O$, far to the left on 
the scale of the figure.  There are two more flows inside 
the tricusp volume than outside.  The tricusp boundary is 
therefore the location of a caustic surface.  We study 
the effects of the gravitational field of a caustic 
surface on the Solar System when the Sun passes through 
it, as for example at one of the double-sided arrows shown.}
\end{center}
\label{fig:tricusp}
\end{figure}

\begin{figure}
\begin{center}
\includegraphics[height=110mm]{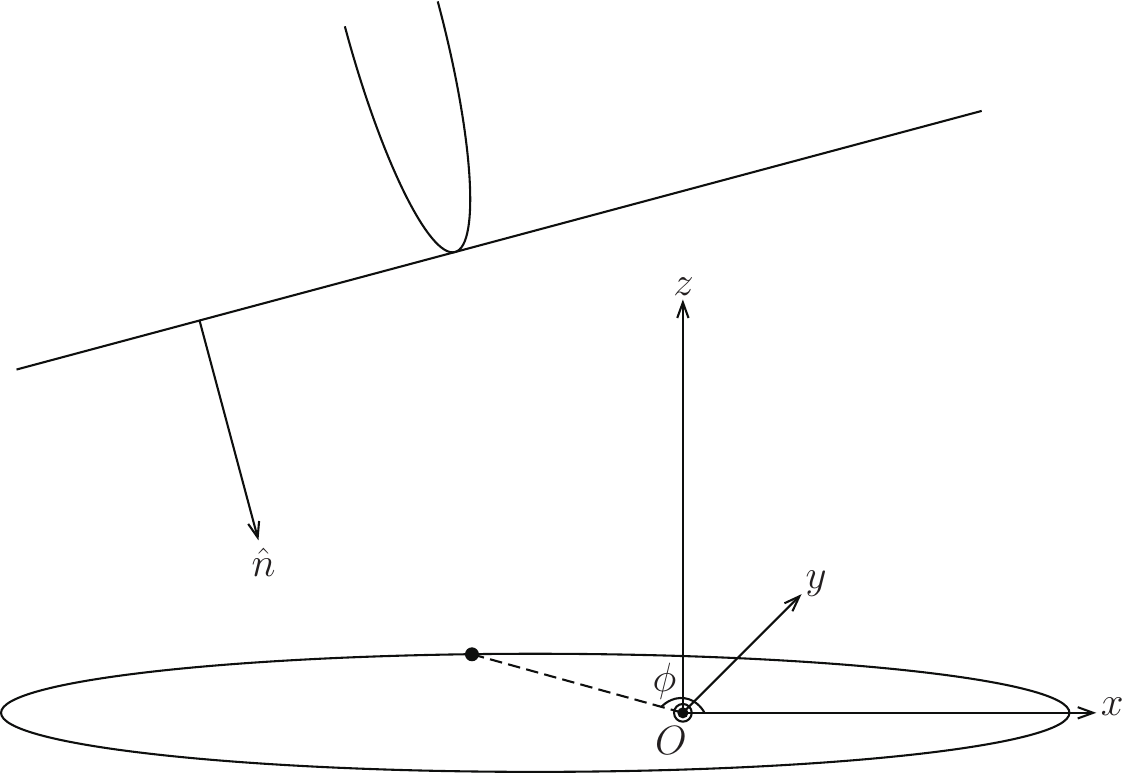}
\vspace{0.3in}
\caption{Schematic drawing of a caustic surface approaching 
the Solar System.   $\hat{n}$ is the normal to the caustic
surface in the direction away from the two extra dark matter
flows.   The Sun is at the origin of the coordinate system.
The ellipse is the initial orbit of a comet.  The comet is 
indicated by the black dot.}
\end{center}
\label{fig:setup}
\end{figure}

\begin{figure}
\begin{center}
\subfigure[]{
\includegraphics[height=55mm]{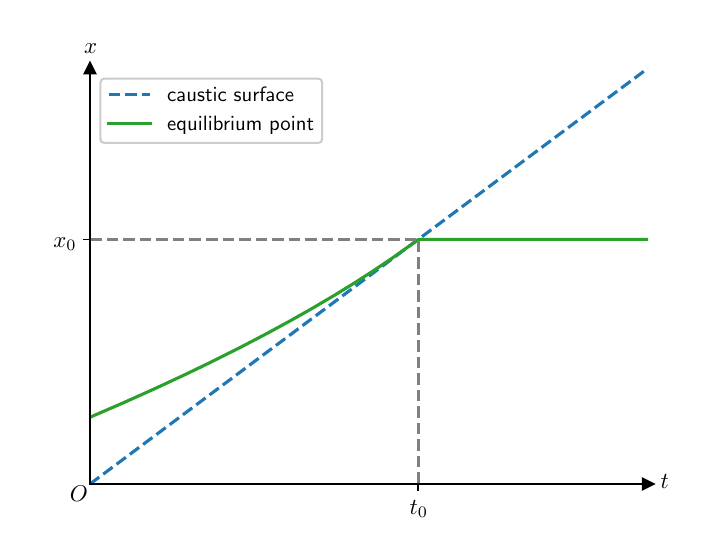}
}
\subfigure[]{
\includegraphics[height=55mm]{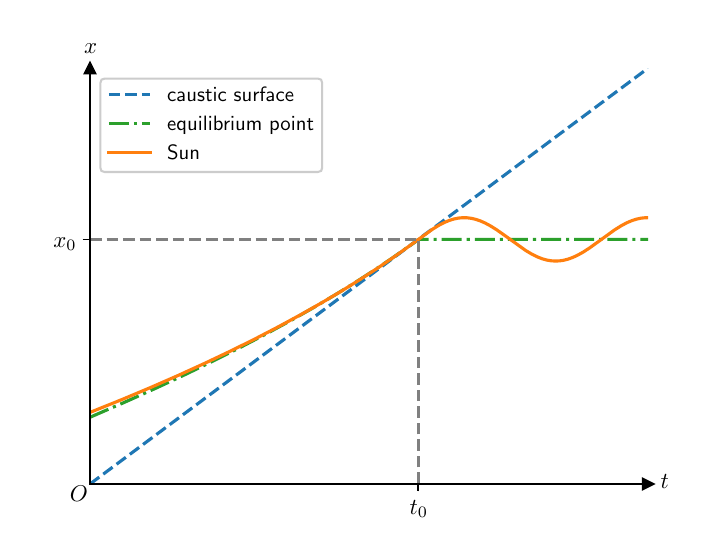}
}
\subfigure[]{
\includegraphics[height=55mm]{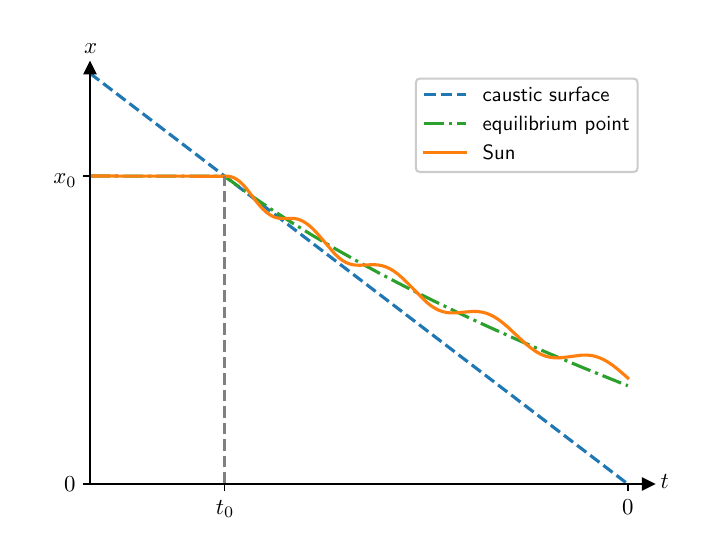}
}
\subfigure[]{
\includegraphics[height=55mm]{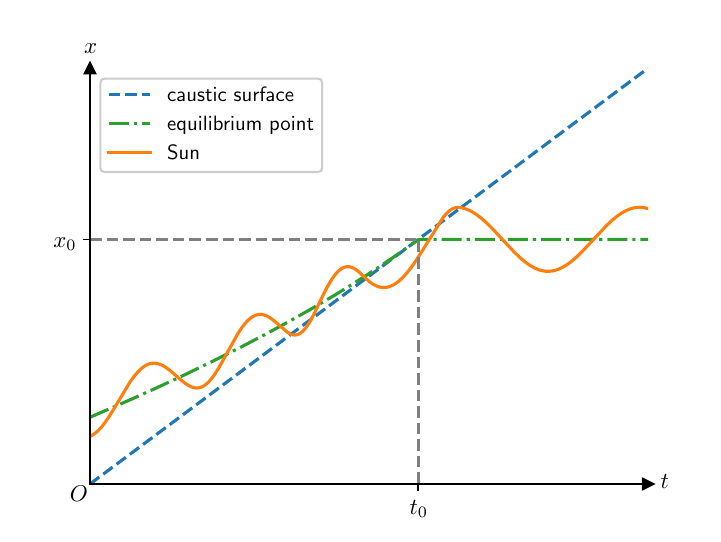}
}
\caption{(a) Motion of the Sun's equilibrium 
position as a caustic passes by.  (b) Motion 
of the Sun as a caustic passes by in case 
the Sun is initially on the side of the 
caustic surface with two extra flows, at
and moving with its equilibrium position. 
(c) Motion of the Sun as a caustic passes 
by in case the Sun is initially on the side 
of the caustic surface without the two extra 
flows, at rest at its equilibrium position.  
$v_c < 0$ here, whereas $v_c > 0$  in (a), 
(b) and (d). (d) Motion of the Sun as a caustic 
passes by in case it is initially oscillating 
about its equilibrium position on the side of 
the caustic surface with two extra 
flows.}
\end{center}
\label{fig:sun}
\end{figure}

\begin{figure}
\begin{center}
\includegraphics[height=110mm]{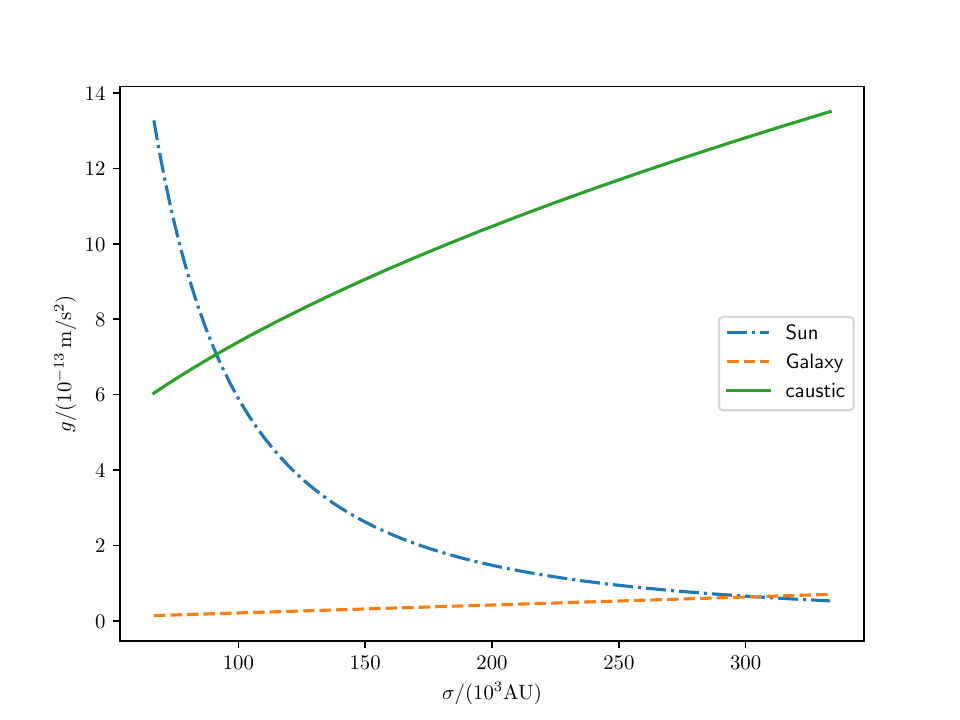}
\vspace{0.3in}
\caption{Accelerations due to the gravitational field of the 
Sun, the tidal field of the Galaxy and the tidal field  of 
the caustic, as functions of the distance between the Sun and
a test particle.  The Sun is assumed to be at the caustic 
surface and the test particle at a distance $\sigma$ from the 
Sun in the direction $- \hat{n}$ defined in Fig.~2.  The 
fold coefficient of the caustic is assumed to be 
$A = 2 \times 10^{-3}$gr/cm$^2\sqrt{\rm kpc}$.   The 
value of $\omega$ characterizing the Galactic tidal field 
is assumed to be $\omega_r$ given in Eq.~(\ref{radom}).} 
\end{center}
\label{fig:forces}
\end{figure}

\begin{figure}
\begin{center}
\includegraphics[height=110mm]{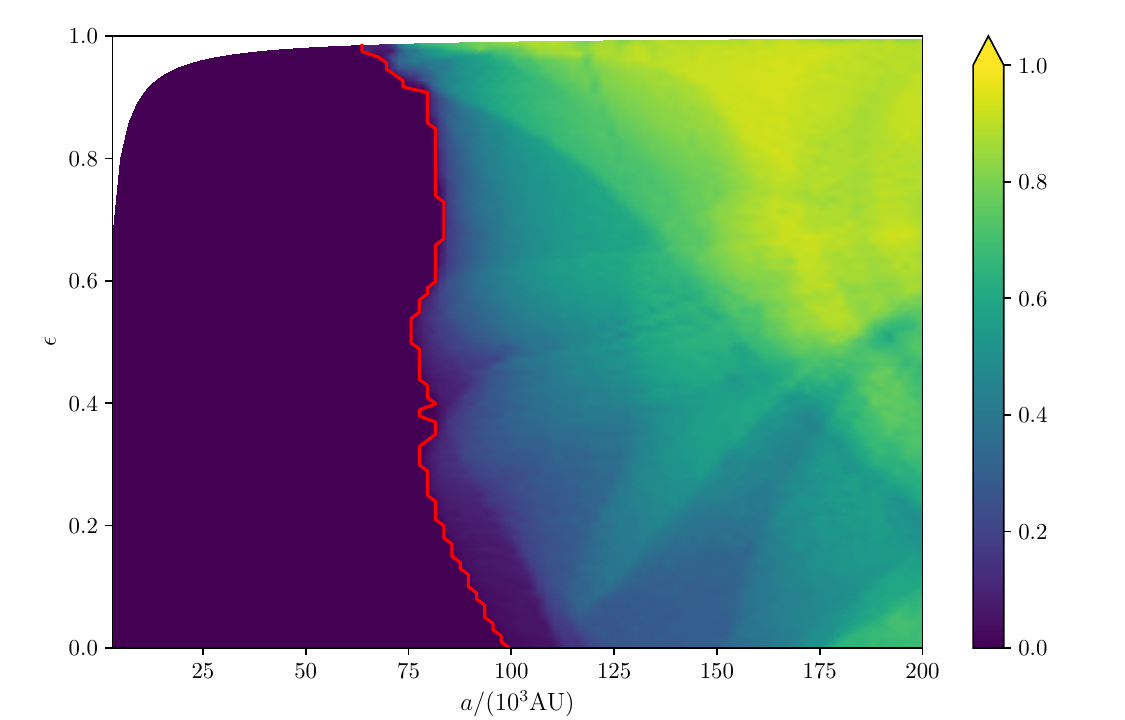}
\vspace{0.3in}
\caption{Probability for a comet to escape the Solar System 
as a function of the semimajor axis $a$ and eccentricity
$\epsilon$ of its initial orbit for the case $A = 2\times
10^{-3}~{\rm gr/cm^2 \sqrt{kpc}}$, $v_c =$ 1 km/s, the Sun 
is initially on the side of the caustic surface with two 
extra flows with $v_\odot$ = 0, $\omega$ = 1/27~Myr, 
$\theta = 90^\circ$ and $\phi_c = 180^\circ$. To the 
left of the red line the escape probability vanishes.}
\end{center}
\label{fig:st_esc}
\end{figure}

\begin{figure}
\begin{center}
\includegraphics[height=110mm]{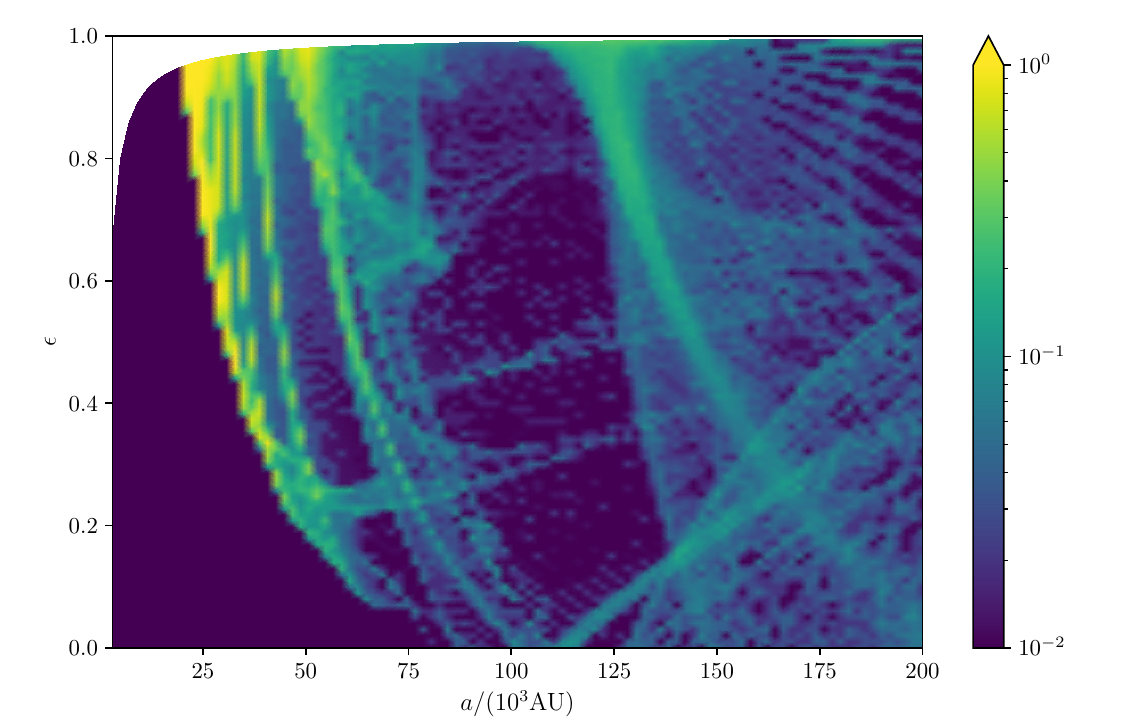}
\vspace{0.3in}
\caption{Probability for a comet to fall within 50 A.U. 
of the Sun as a function of the semimajor axis $a$ and 
eccentricity $\epsilon$ of its initial orbit for the case 
$A = 2\times 10^{-3}~{\rm gr/cm^2 \sqrt{kpc}}$, $v_c =$ 1 km/s, 
the Sun is initially on the side of the caustic surface with 
two extra flows with $v_\odot$ = 0, $\omega$ = 1/27~Myr,
$\theta = 90^\circ$ and $\phi_c = 180^\circ$.}
\end{center}
\label{fig:st_inf}
\end{figure}

\begin{figure}
\begin{center}
\subfigure[]{
\includegraphics[height=50mm]{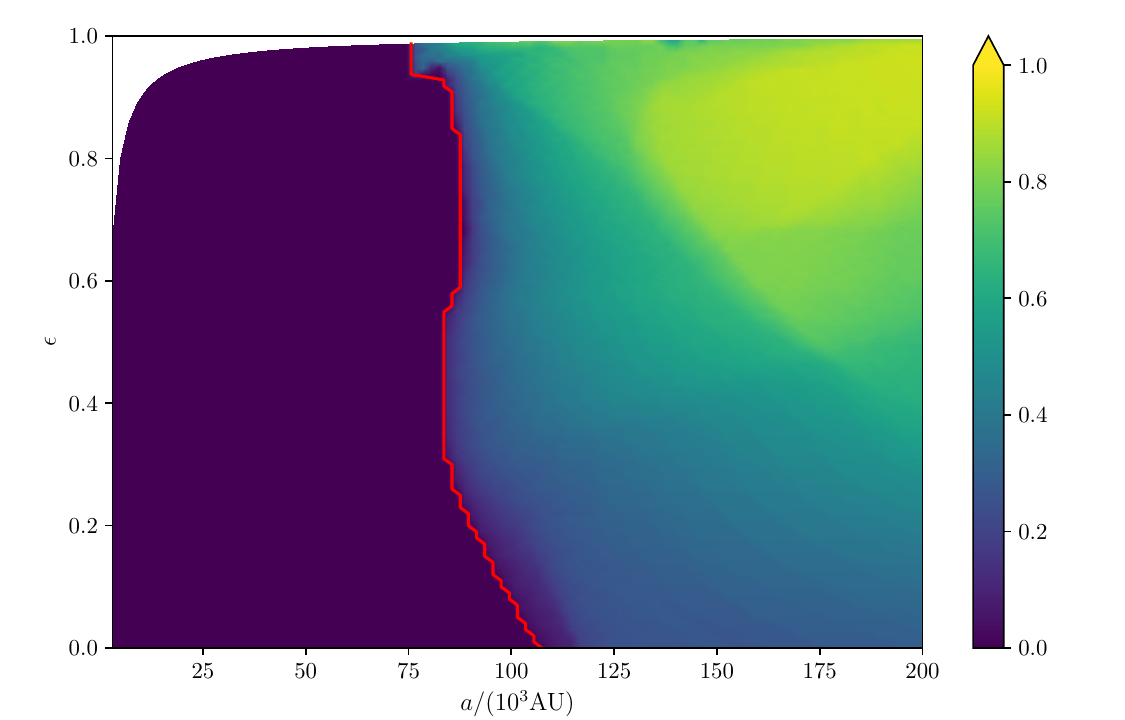}
}
\subfigure[]{
\includegraphics[height=50mm]{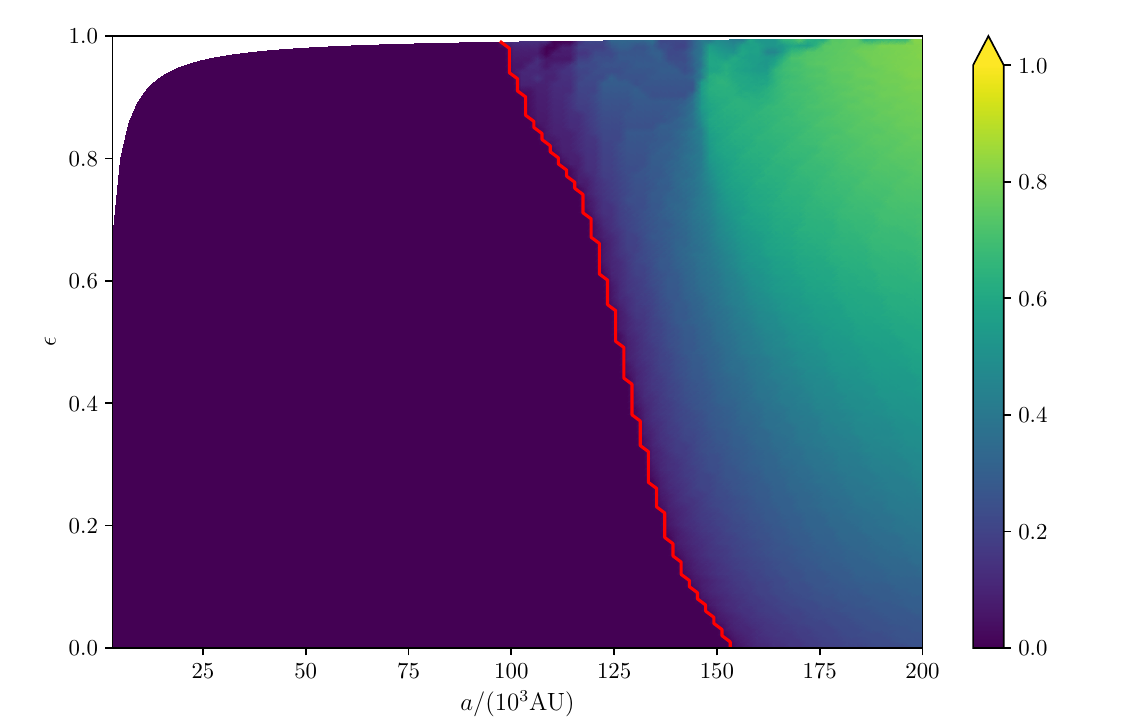}
}
\caption{Same as Fig.~5 except (a) $\theta =
60^\circ$ and (b) $\theta = 30^\circ$.}
\end{center}
\label{fig:esc_th}
\end{figure}

\begin{figure}
\begin{center}
\subfigure[]{
\includegraphics[height=50mm]{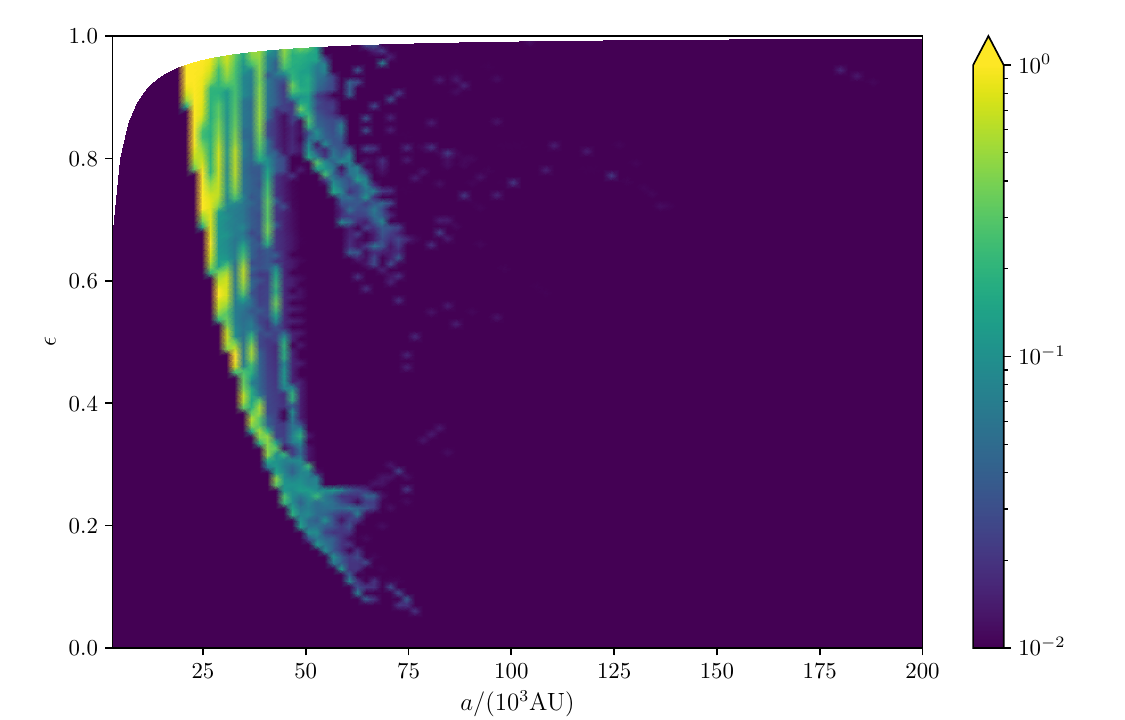}
}
\subfigure[]{
\includegraphics[height=50mm]{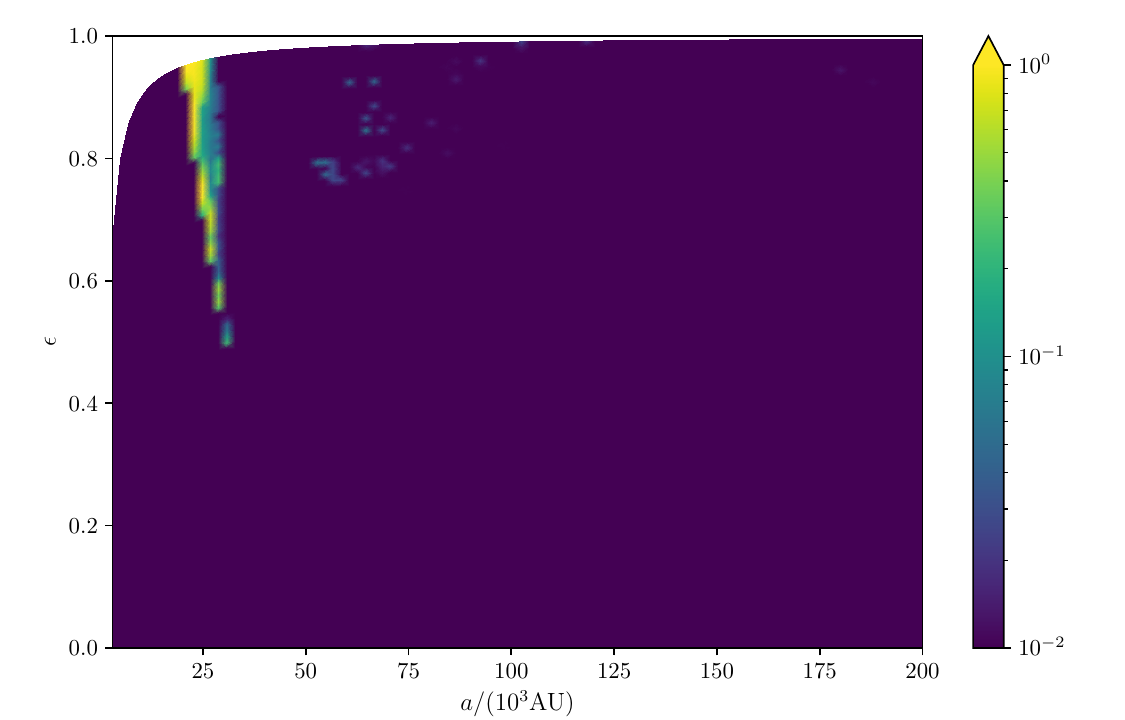}
}
\caption{Same as Fig.~6 except (a) $\theta =
88^\circ$ and (b) $\theta = 87^\circ$.}
\end{center}
\label{fig:inf_th}
\end{figure}

\begin{figure}
\begin{center}
\subfigure[]{
\includegraphics[height=50mm]{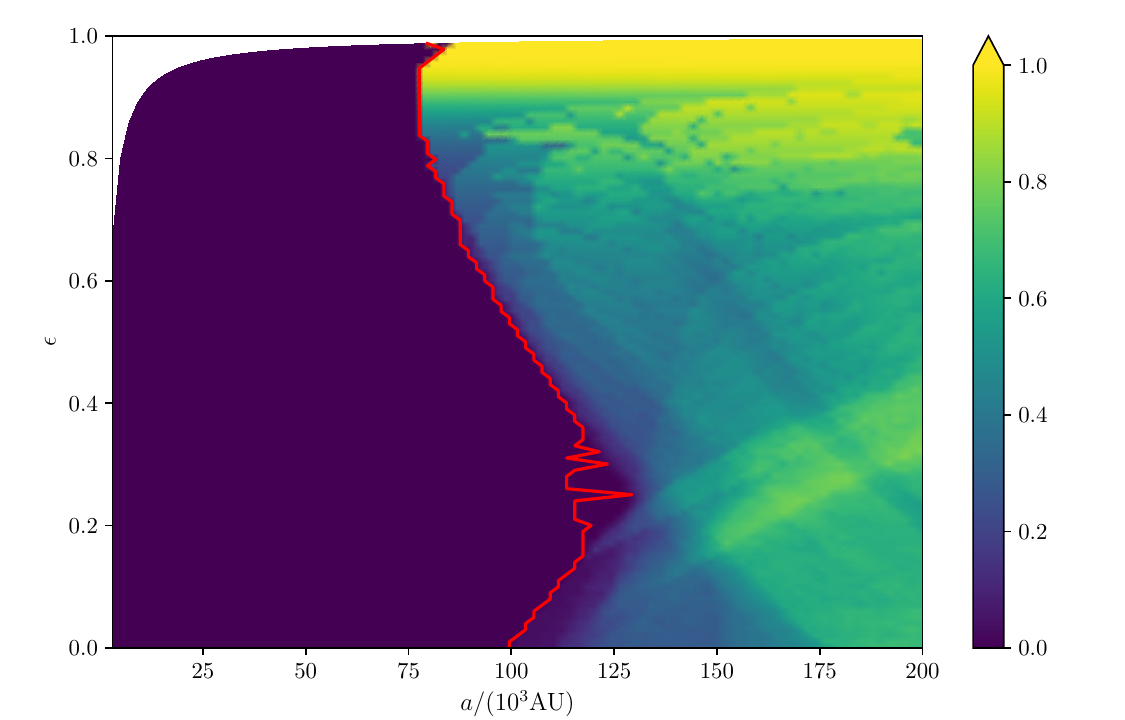}
}
\subfigure[]{
\includegraphics[height=50mm]{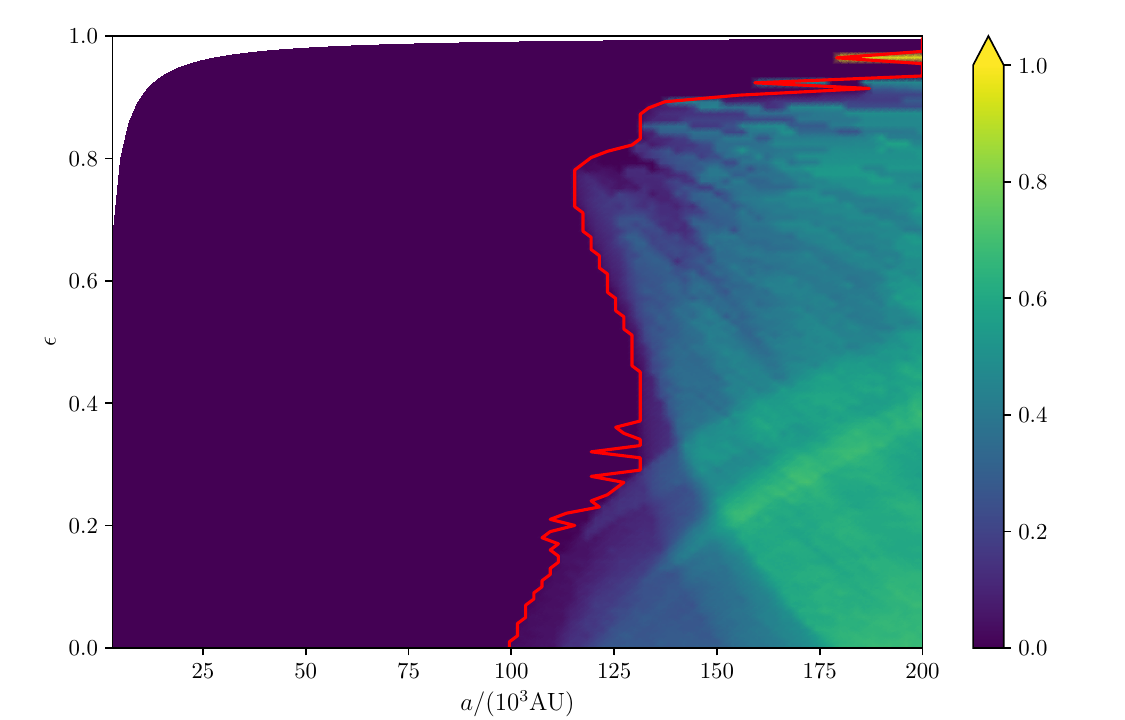}
}
\caption{Same as Fig.~5 except (a) $\phi_c =
0^\circ$ and (b) $\phi_c= 90^\circ$.}
\end{center}
\label{fig:esc_phc}
\end{figure}

\begin{figure}
\begin{center}
\subfigure[]{
\includegraphics[height=50mm]{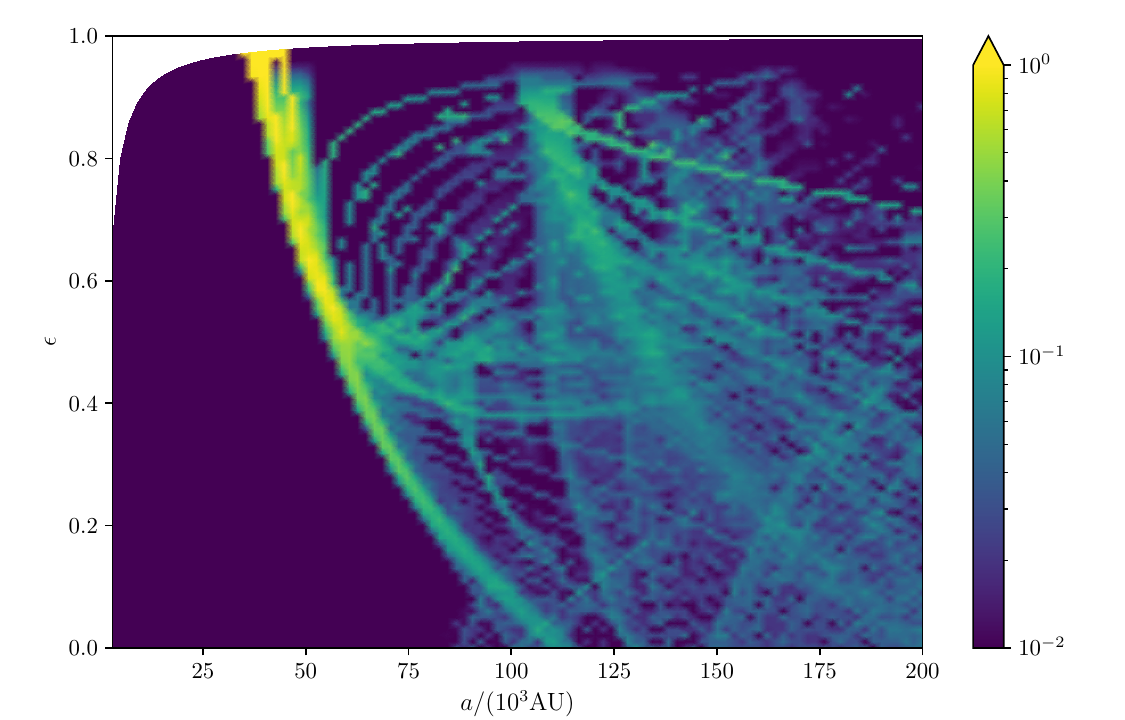}
}
\subfigure[]{
\includegraphics[height=50mm]{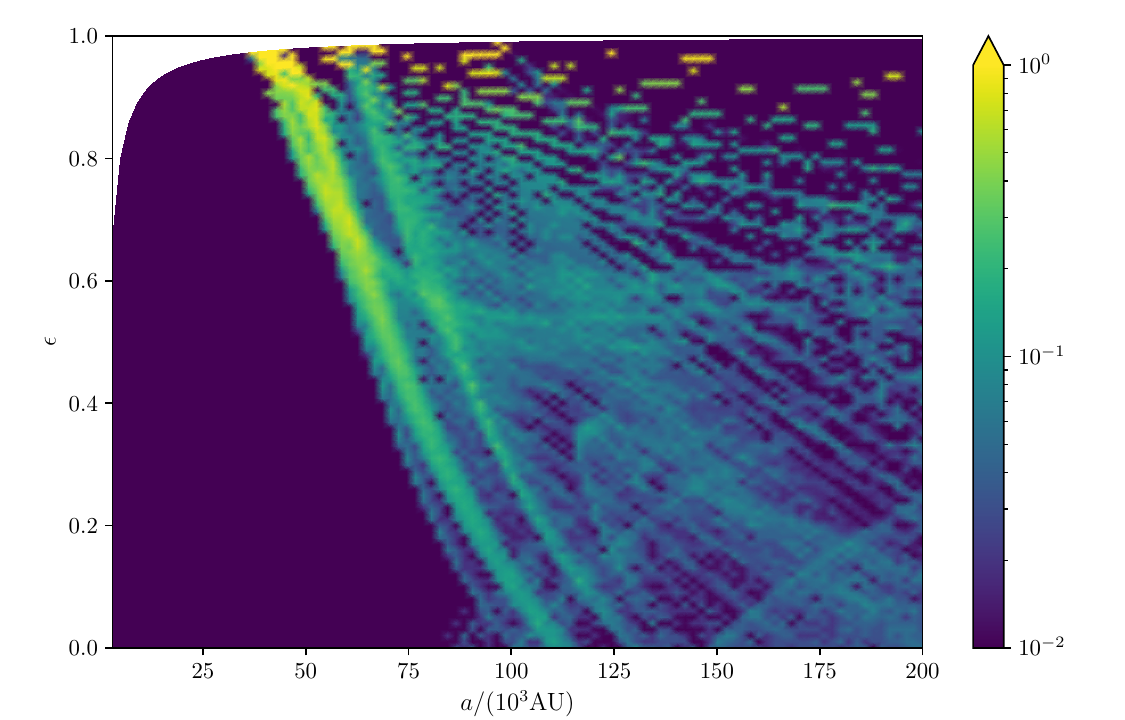}
}
\caption{Same as Fig.~6 except (a) $\phi_c=
0^\circ$ and (b) $\phi_c = 90^\circ$.}
\end{center}
\label{fig:inf_phc}
\end{figure}

\begin{figure}
\begin{center}
\subfigure[]{
\includegraphics[height=50mm]{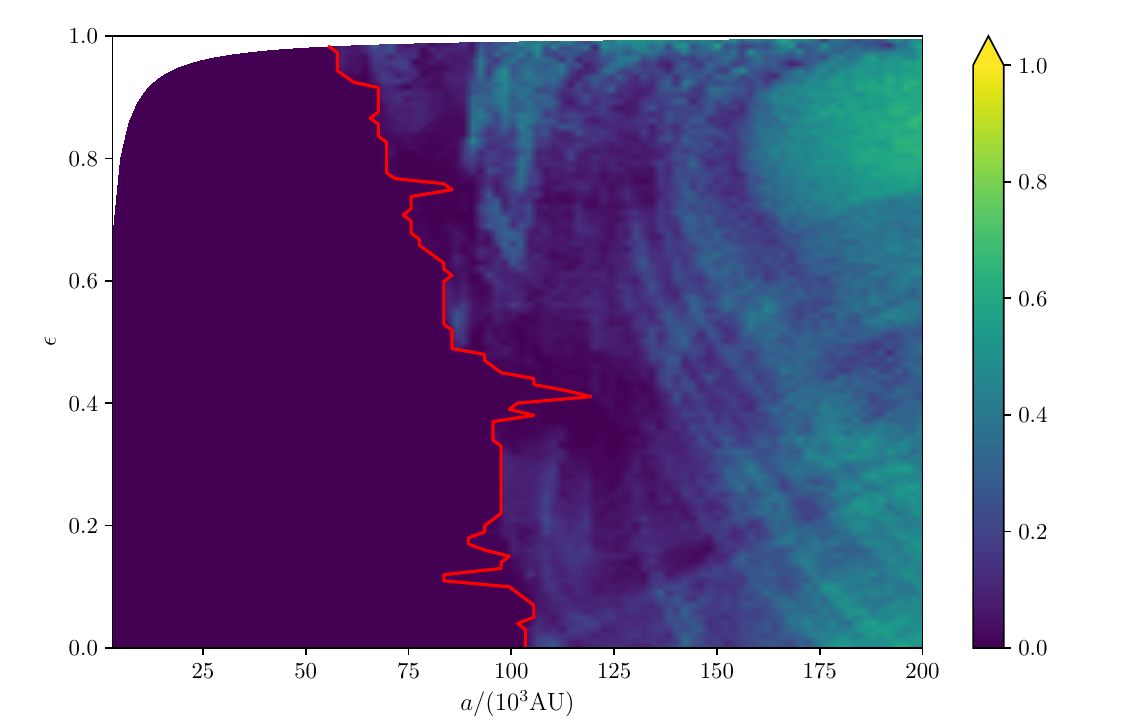}
}
\subfigure[]{
\includegraphics[height=50mm]{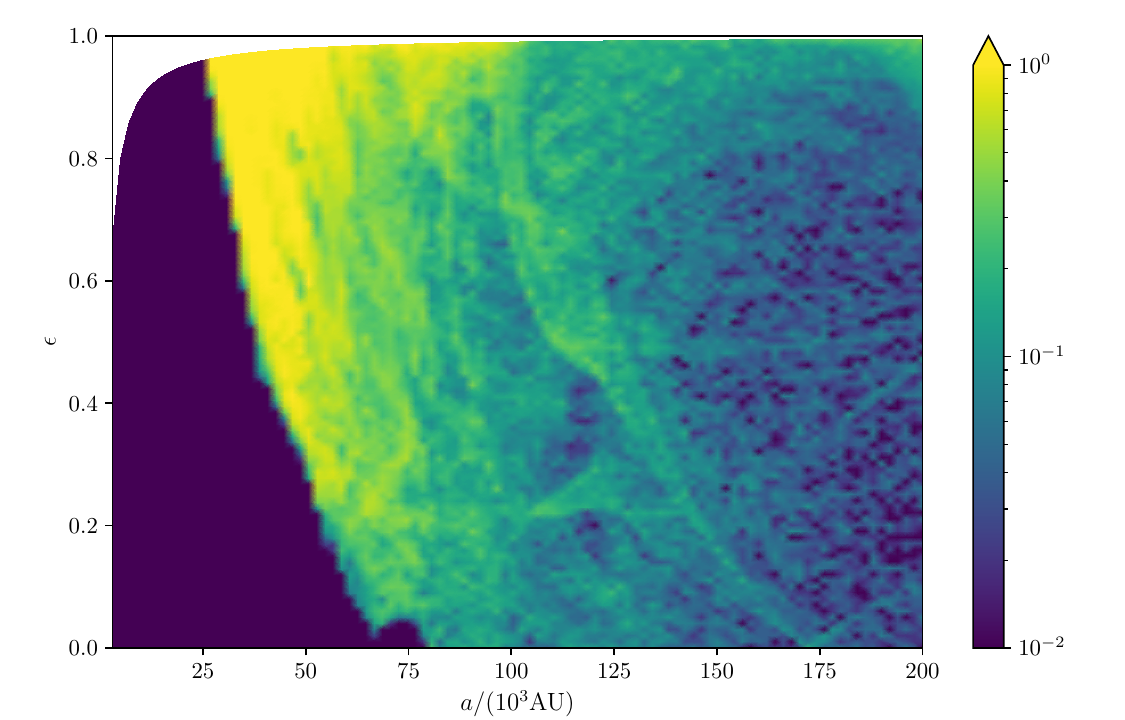}
}
\caption{Here the Sun initially oscillates 
about its equilibrium position, as in Fig.~3(d).
(a) Changes to Fig.~5.  (b) Changes to Fig.~6.} 
\end{center}
\label{fig:osc}
\end{figure}

\begin{figure}
\begin{center}
\subfigure[]{
\includegraphics[height=50mm]{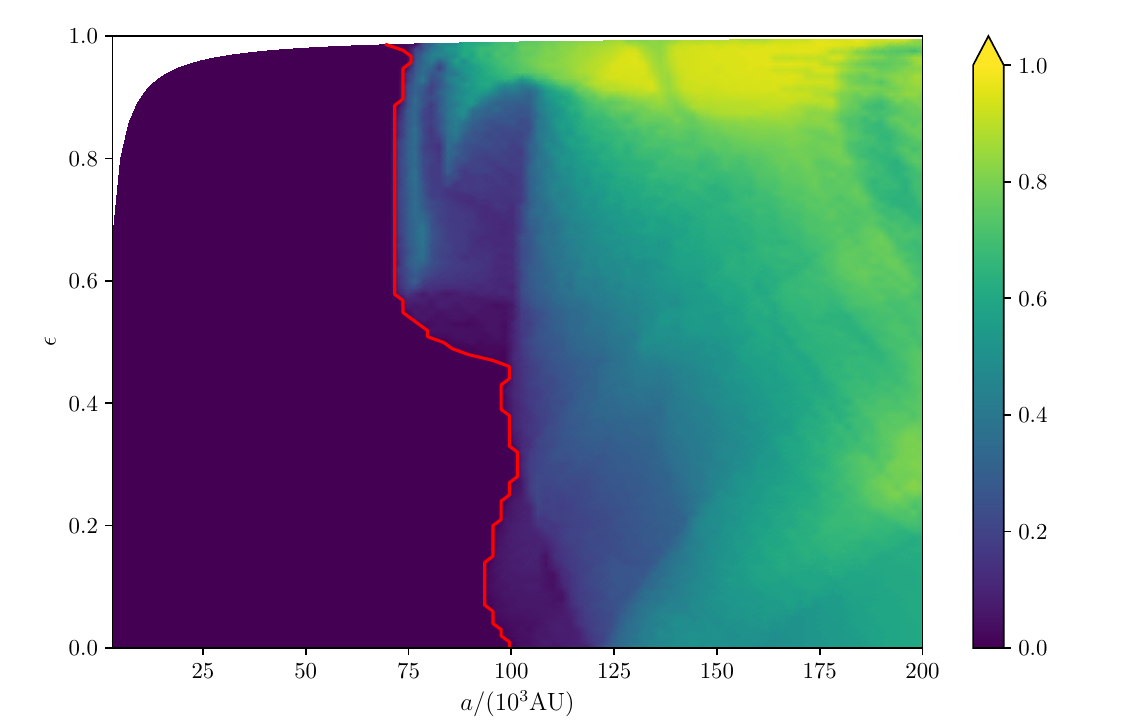}
}
\subfigure[]{
\includegraphics[height=50mm]{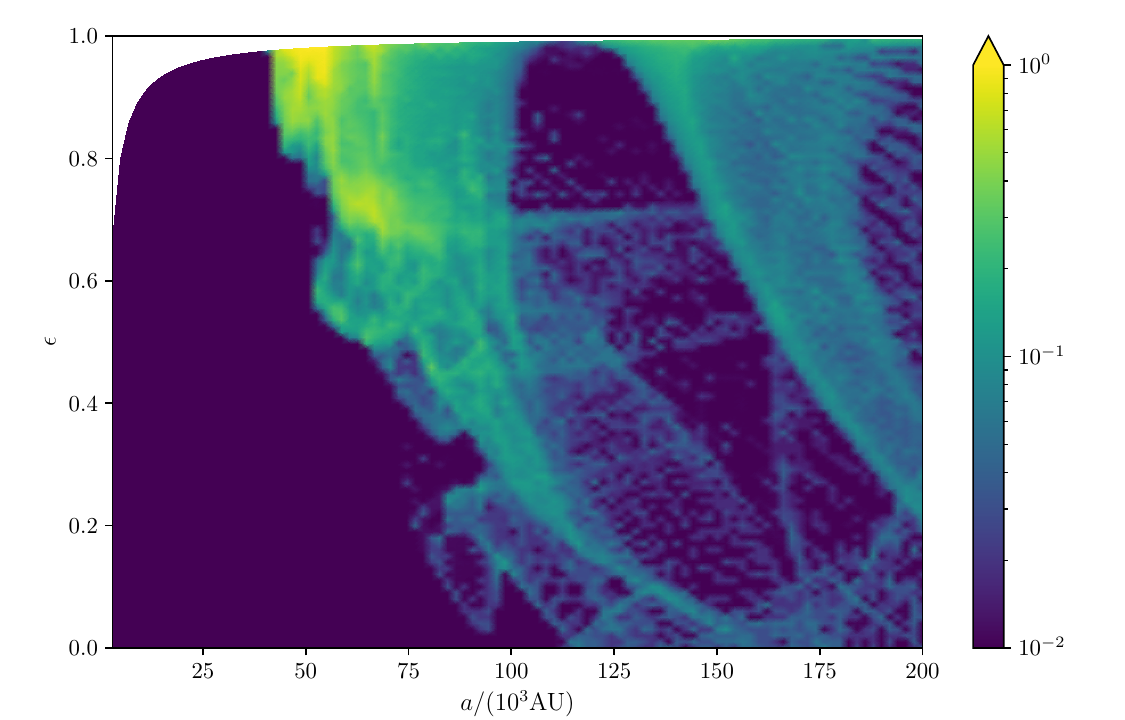}
}
\caption{Here the Sun is initially at rest 
on the side of the caustic surface without 
the two extra flows, as in Fig.~3(c).  (a)
Changes to Fig.~5. (b) Changes to Fig.~6.}
\end{center}
\label{fig:rev}
\end{figure}

\begin{figure}
\begin{center}
\subfigure[]{
\includegraphics[height=50mm]{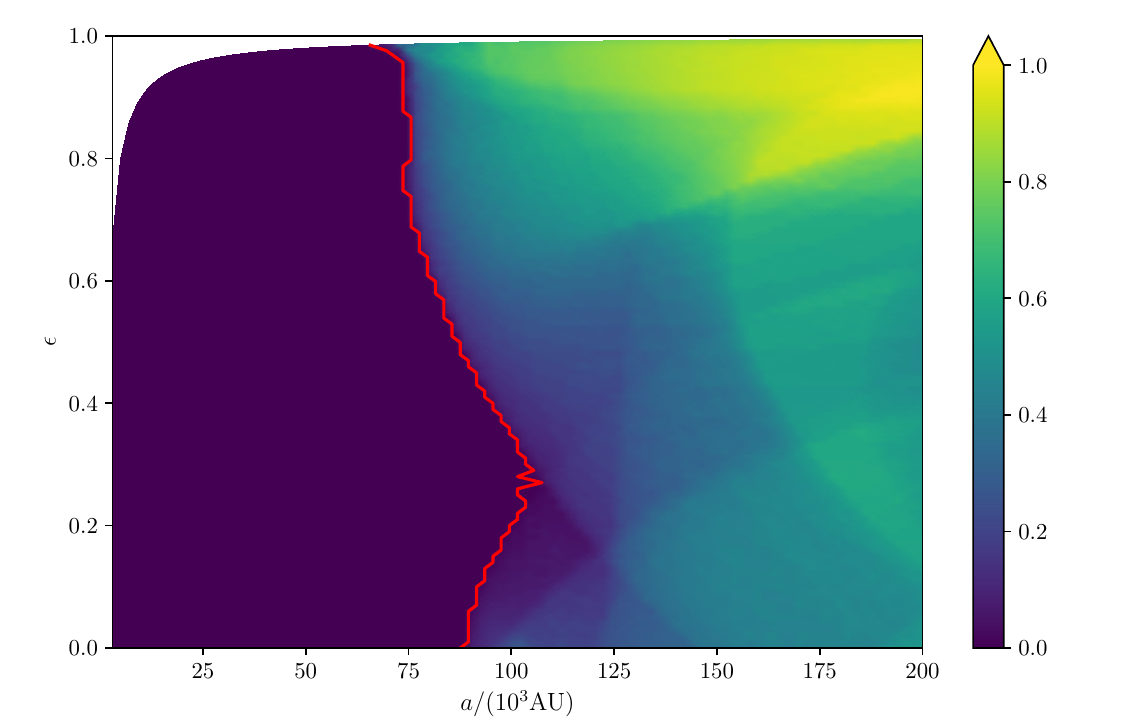}
}
\subfigure[]{
\includegraphics[height=50mm]{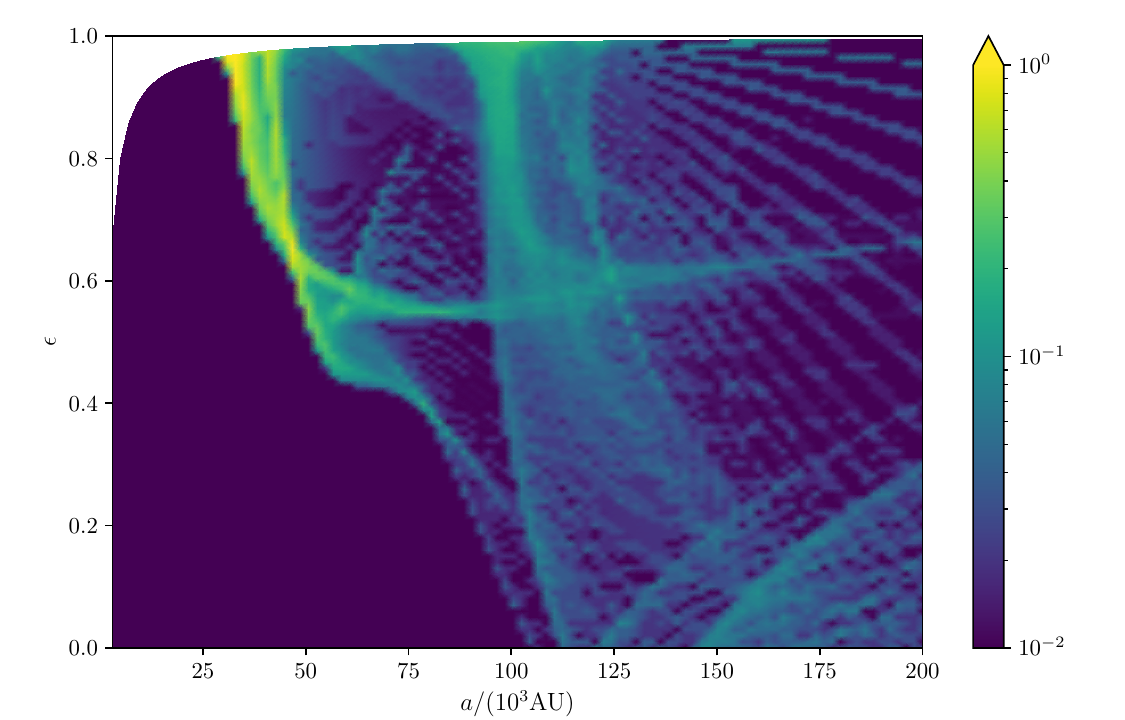}
}
\caption{(a) Same as Fig.~5 except $\omega$
= 1/11~Myr. (b) Same as Fig.~6 except 
$\omega$ = 1/11~Myr.}
\end{center}
\label{fig:ome}
\end{figure}

\begin{figure}
\begin{center}
\subfigure[]{
\includegraphics[height=50mm]{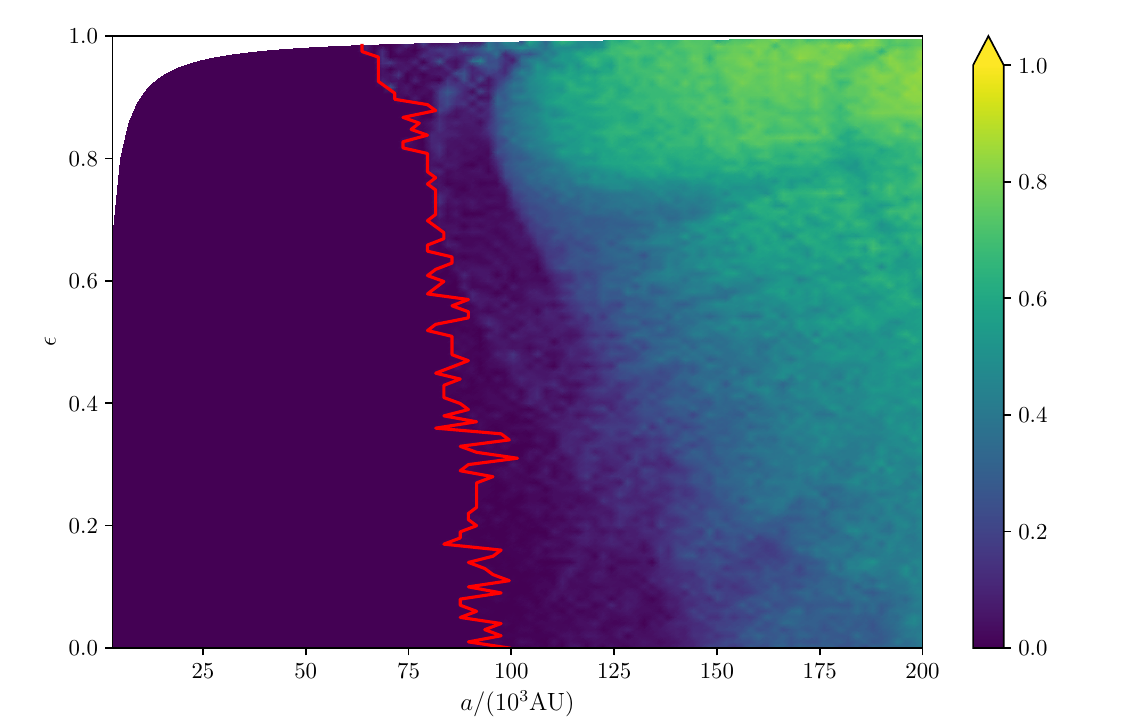}
}
\subfigure[]{
\includegraphics[height=50mm]{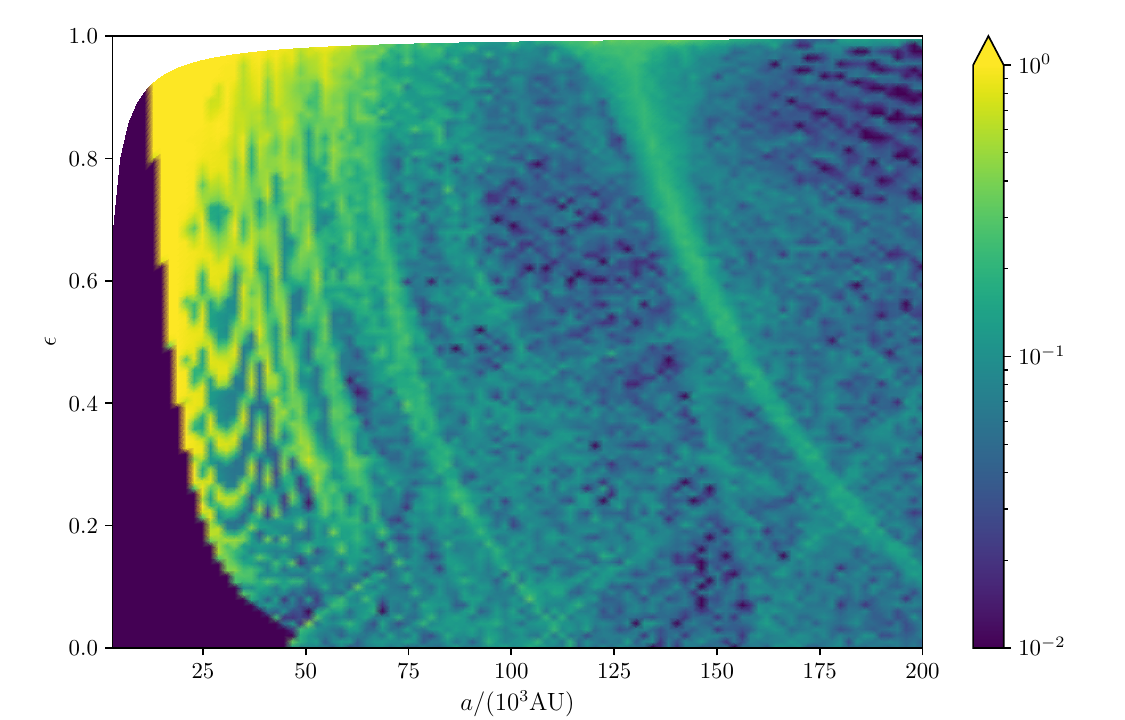}
}
\subfigure[]{
\includegraphics[height=50mm]{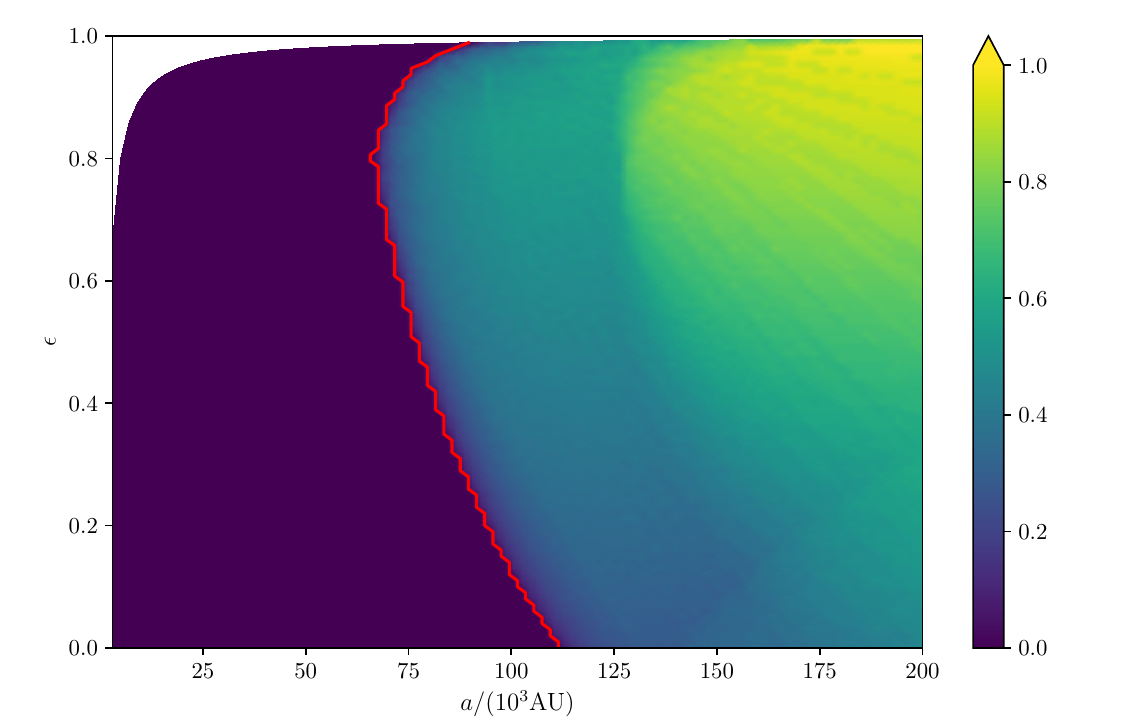}
}
\subfigure[]{
\includegraphics[height=50mm]{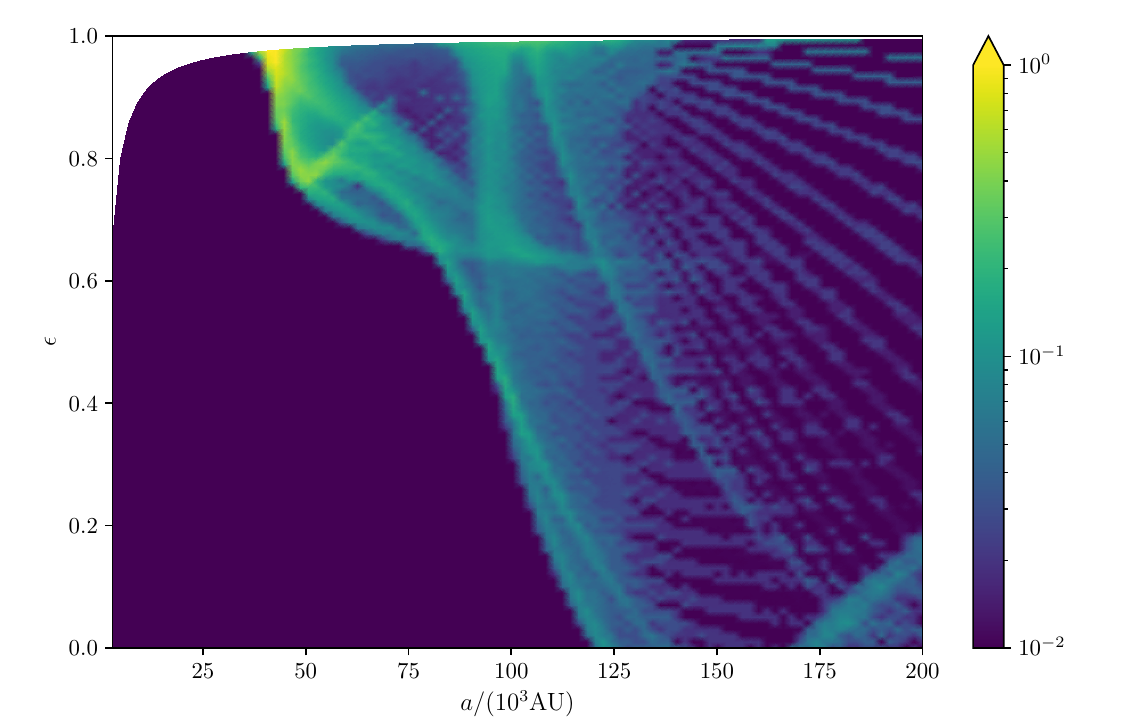}
}
\caption{(a) Same as Fig.~5 except 
$v_c$ = 0.3 km/s.  (b) Same as Fig.~6 
except $v_c$ = 0.3 km/s.  \\ (c) Same as 
Fig.~5 except $v_c$ = 10 km/s.  (d) 
Same as Fig.~6 except $v_c$ = 10 km/s.}
\end{center}
\label{fig:vc}
\end{figure}

\begin{figure}
\begin{center}
\subfigure[]{
\includegraphics[height=50mm]{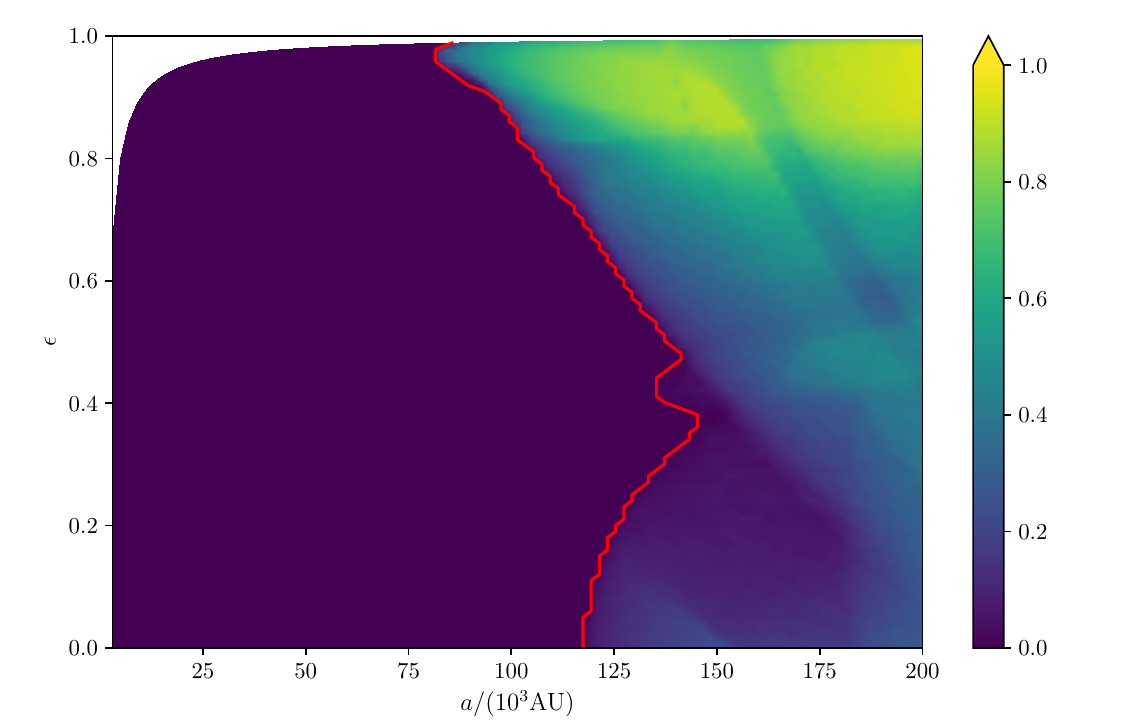}
}
\subfigure[]{
\includegraphics[height=50mm]{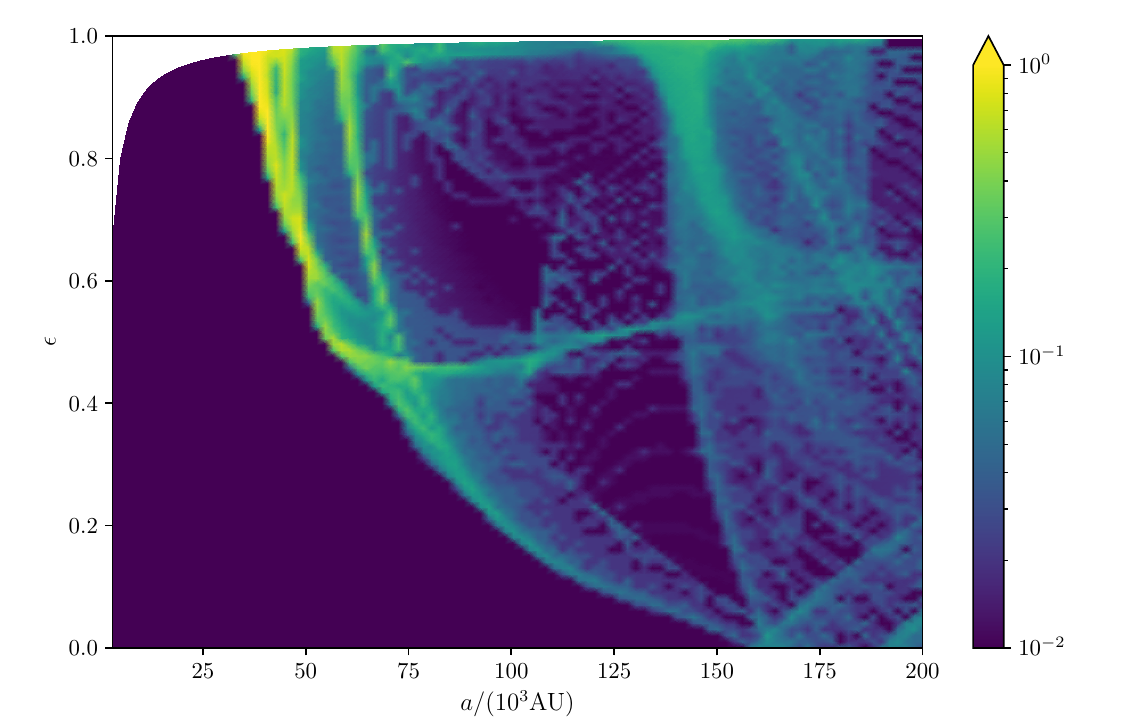}
}
\subfigure[]{
\includegraphics[height=50mm]{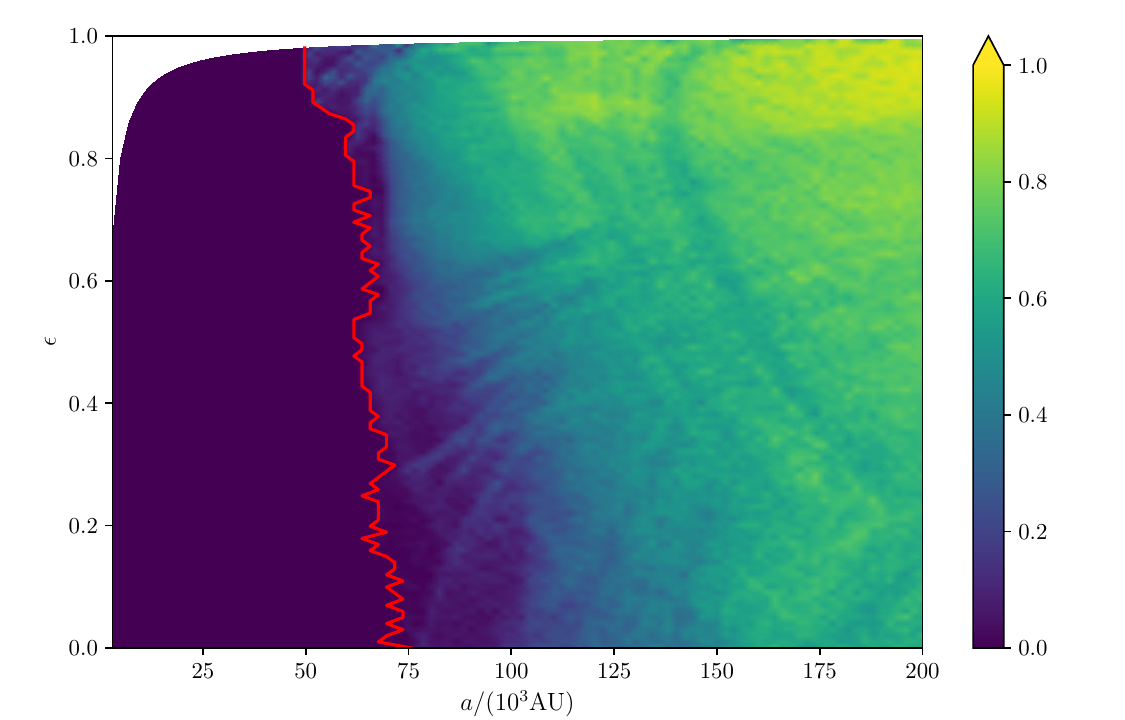}
}
\subfigure[]{
\includegraphics[height=50mm]{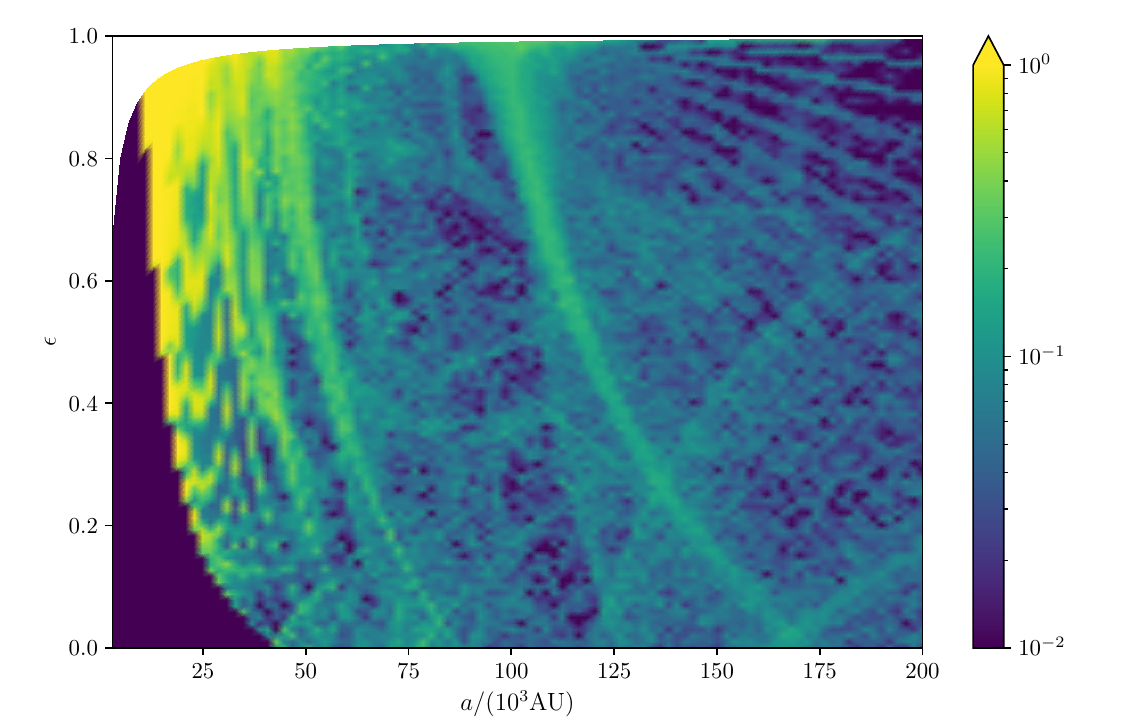}
}
\caption{(a) Same as Fig.~5 except
$A = 1.0\times 10^{-3}\,{\rm gr/cm^2 \sqrt{kpc}}$.  
(b) Same as Fig.~6 except 
$A = 1.0\times 10^{-3}\,{\rm gr/cm^2 \sqrt{kpc}}$.
(c) Same as Fig.~5 except 
$A = 4.0\times 10^{-3}\,{\rm gr/cm^2 \sqrt{kpc}}$.
(d) Same as Fig.~6 except 
$A = 4.0\times 10^{-3}\,{\rm gr/cm^2 \sqrt{kpc}}$.}
\end{center}
\label{fig:A}
\end{figure}

\begin{figure}
\begin{center}
\subfigure[]{
\includegraphics[height=50mm]{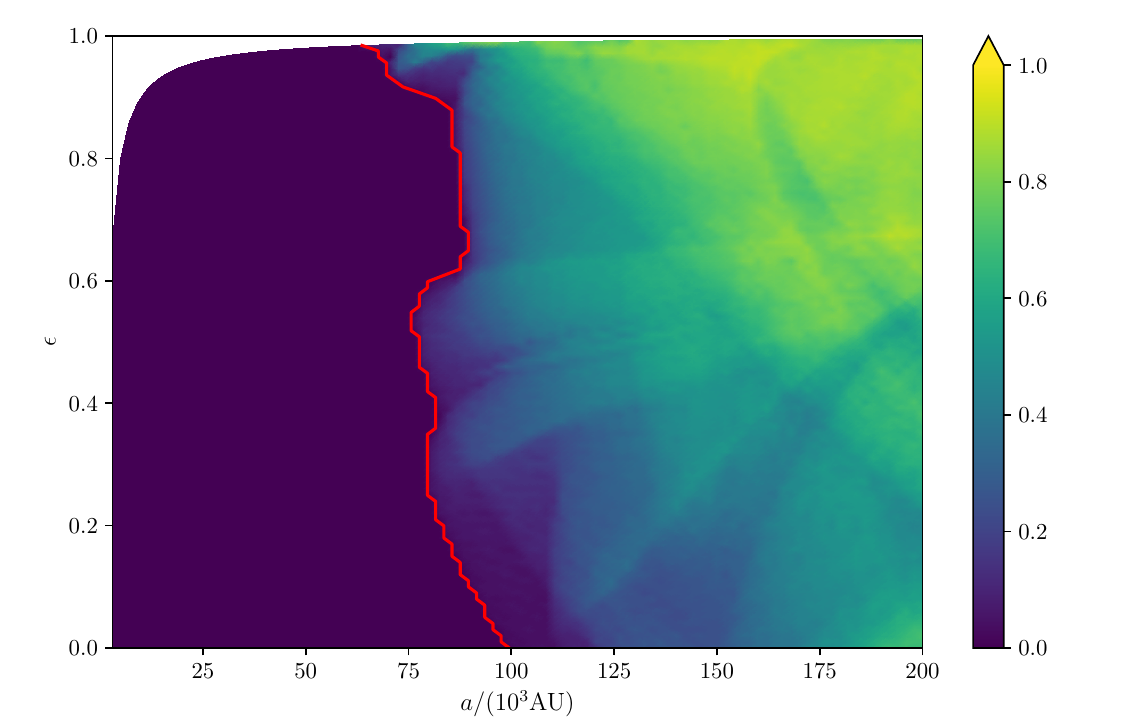}
}
\subfigure[]{
\includegraphics[height=50mm]{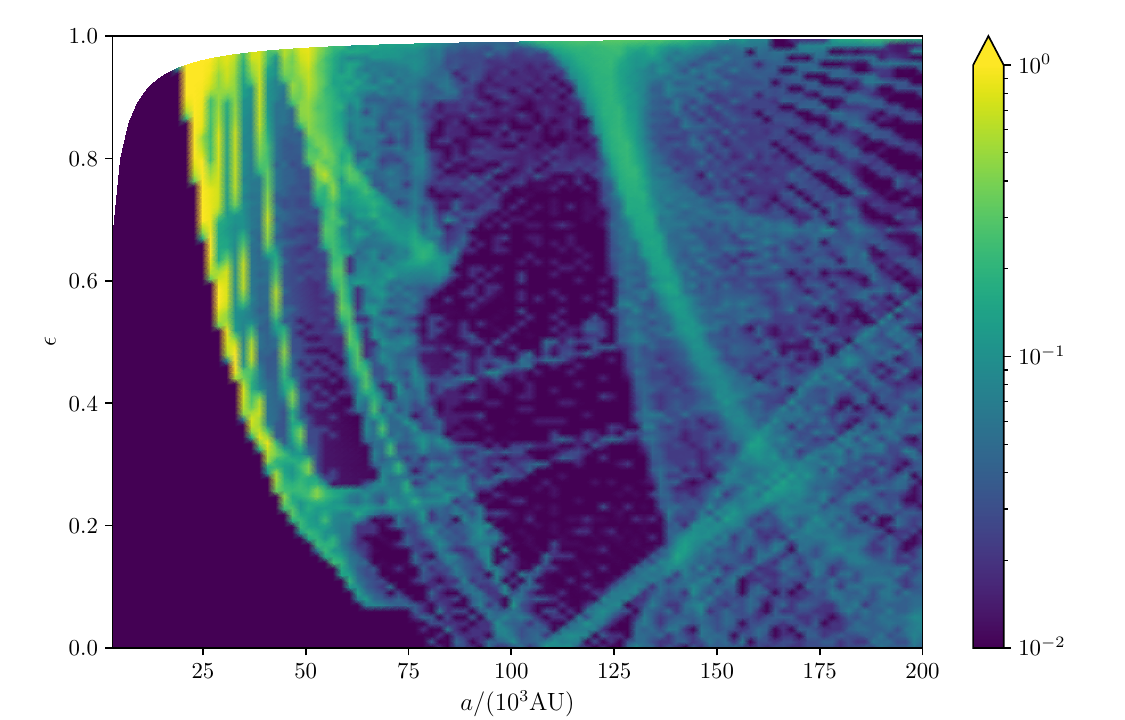}
}
\caption{Here the caustic is divided into 
seven equal parallel caustics spread over 
a transverse distance of 2 pc.  (a) Changes 
to Fig.~5.  (b) Changes to Fig.~6.}
\end{center}
\label{fig:n7}
\end{figure}

\begin{figure}
\begin{center}
\includegraphics[height=110mm]{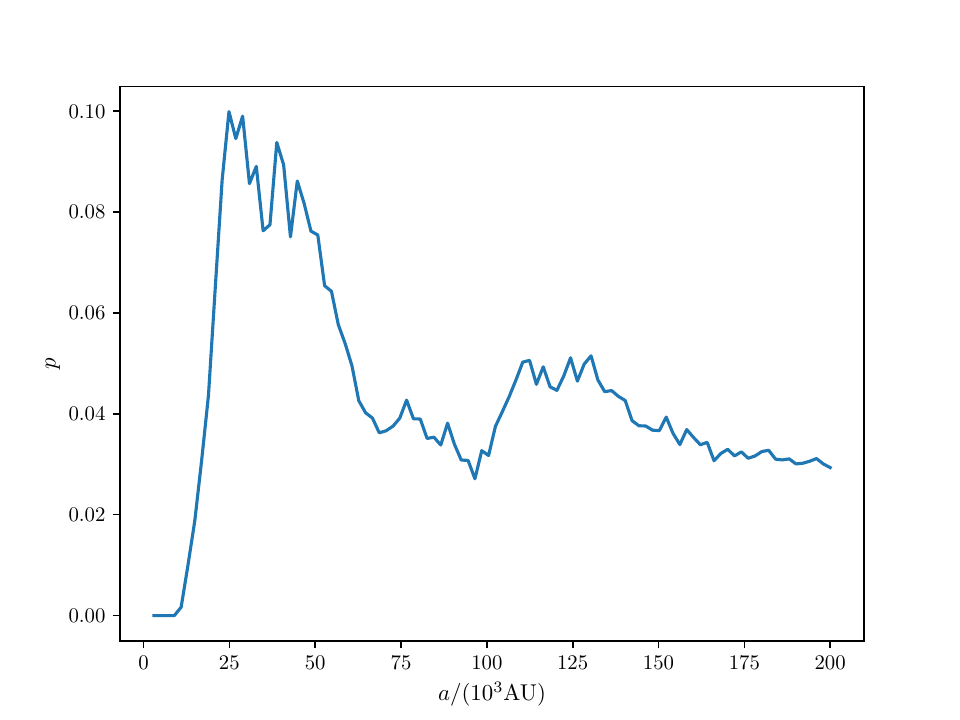}
\vspace{0.3in}
\caption{Average probability for a comet to 
fall within 50 A.U.of the Sun as a function 
of its initial semimajor axis $a$ for the case
$A = 2\times 10^{-3}~{\rm gr/cm^2 \sqrt{kpc}}$, 
$v_c =$ 1 km/s, the Sun is initially on the 
side of the caustic surface with two extra 
flows with $v_\odot$ = 0, $\omega$ = 1/27 Myr, and
$\theta = 90^\circ$.  The probability is 
averaged over $\epsilon$, $\phi_0$, and $\phi_c$
as described in the text.}
\end{center}
\label{fig:inf_prob}
\end{figure}

\clearpage

\end{document}